\documentclass[a4paper,12pt]{article}
\pdfoutput=1
\usepackage{feynmp-auto,expdlist}
\usepackage{amsmath, amsfonts, amssymb}
\usepackage{graphicx}
\usepackage{enumerate}
\usepackage{hyperref}
\usepackage{latexsym}
\usepackage{hepnicenames}
\usepackage{enumerate}
\usepackage{soul}
\usepackage[normalem]{ulem}
\usepackage{bbold}
\usepackage{wasysym}
\usepackage{makecell}

\font\tenrsfs=rsfs10 at 12pt
\font\sevenrsfs=rsfs7
\font\fiversfs=rsfs5
\newfam\rsfsfam
\textfont\rsfsfam=\tenrsfs
\scriptfont\rsfsfam=\sevenrsfs
\scriptscriptfont\rsfsfam=\fiversfs

\numberwithin{equation}{section}

\usepackage{mathrsfs}
\usepackage{braket}
\usepackage{titling}
\usepackage{amsmath}
\usepackage{slashed}
\usepackage{amssymb}
\usepackage{epsfig}
\usepackage{graphicx}
\usepackage{color}
\usepackage{rotating}
\usepackage{hyperref}
\usepackage[margin=1.in]{geometry}
\usepackage[table,xcdraw,dvipsnames]{xcolor}
\usepackage[compress,numbers,sort]{natbib}
\usepackage{colortbl}
\usepackage{pdflscape}
\usepackage{color}
\usepackage{mathtools}
\usepackage{colortbl}
\definecolor{Gray}{gray}{0.95}
\definecolor{RGray}{gray}{0.85}
\definecolor{CGray}{gray}{0.92}

\newcommand{\A}{{\cal A}}

\newcommand{\X}{{\cal X}}
\newcommand{\G}{{\cal G}}
\renewcommand{\P}{{\cal P}}
\newcommand{\C}{{\cal C}}

\newcommand{\SU}{{\rm SU}}
\newcommand{\SO}{{\rm SO}}

\newcommand{\U}{{\rm U}}
\newcommand{\Z}{{\cal Z}}
\newcommand{\V}{{\cal V}}
\newcommand{\M}{{\cal M}}

\definecolor{nicered}{rgb}{0.7,0.1,0.1}
\definecolor{nicegreen}{rgb}{0.1,0.5,0.1}
\definecolor{red}{rgb}{1.0, 0, 0}
\definecolor{niceblue}{rgb}{0,0,0.8}
\definecolor{red}{rgb}{1.0, 0, 0}
\hypersetup{colorlinks,bookmarksopen,bookmarksnumbered,
linkcolor=blus,pdfstartview=FitH,urlcolor=rossos,citecolor=verde}
\allowdisplaybreaks

\definecolor{rosso}{cmyk}{0,1,1,0.4}
\definecolor{rossos}{cmyk}{0,1,1,0.55}
\definecolor{rossoc}{cmyk}{0,1,1,0.2}
\definecolor{blu}{cmyk}{1,1,0,0.3}
\definecolor{blus}{cmyk}{1,1,0,0.6}
\definecolor{bluc}{cmyk}{1,1,0,0.1}
\definecolor{verde}{cmyk}{0.92,0,0.59,0.25}
\definecolor{verdec}{cmyk}{0.92,0,0.59,0.15}
\definecolor{verdes}{cmyk}{0.92,0,0.59,0.4}

\def\eq#1{{Eq.~(\ref{#1})}}
\def\eqs#1#2{{Eqs.~(\ref{#1})--(\ref{#2})}}
\def\fig#1{{Fig.~\ref{#1}}}

\def\Table#1{{Table~\ref{#1}}}

\def\sect#1{{Sect.~\ref{#1}}}

\def\app#1{{App.~\ref{#1}}}

\def\vev#1{\left\langle #1\right\rangle}

\def\Tr{\mbox{Tr}\,}

\def\diag{\mbox{diag}\,}

\renewcommand{\bar}{\overline}

\newcommand{\beq}{\begin{equation}}
\newcommand{\eeq}{\end{equation}}
\newcommand{\bea}{\begin{eqnarray}}
\newcommand{\eea}{\end{eqnarray}}

\renewcommand{\[}{\left[}
\renewcommand{\]}{\right]}
\renewcommand{\(}{\left(}
\renewcommand{\)}{\right)}

\newcommand{\Q}{\mathcal{Q}}

\renewcommand{\S}{\mathcal{S}}
\renewcommand{\X}{\mathcal{X}}

\begin{document}
\vspace{1.5cm}


\begin{center}
{\Large\LARGE\Huge\bf\color{blus} Pati-Salam Axion}\\[1cm]
{\bf Luca Di Luzio}\\[7mm]

{\it 
DESY, Notkestra\ss e 85, 
D-22607 Hamburg, Germany}\\[1mm]

\vspace{0.5cm}

\begin{quote}\large
I discuss the implementation of the Peccei-Quinn mechanism 
in a minimal realization of the Pati-Salam partial unification 
scheme. The axion mass is shown to be related to the 
Pati-Salam breaking scale and it is predicted 
via a two-loop renormalization group analysis
to be in the window $m_a \in [10^{-11}, \, 3 \times 10^{-7}]$ eV,  
as a function of a sliding Left-Right symmetry breaking scale. 
This parameter space will be fully covered by the late phases 
of the axion Dark Matter experiments ABRACADABRA and CASPEr-Electric. 
A Left-Right symmetry breaking scenario as low as 20 TeV 
is obtained for a Pati-Salam breaking of the order of the reduced Planck mass. 

\end{quote}

\thispagestyle{empty}
\bigskip

\end{center}

\setcounter{footnote}{0}

\newpage
\tableofcontents

\newpage

\section{Introduction}
\label{sec:intro}

A central question of physics beyond the 
Standard Model (SM) is whether there is intermediate-scale physics  
between the electroweak and the Planck scales and how to possibly test it. 
It can be reasonably argued that the SM is an effective field theory, 
valid until some cut-off scale $\Lambda_{\rm SM} \leq M_{\rm Pl} = 1.2 \times 10^{19}$ GeV
and that (disregarding the long-pursued naturalness argument of the electroweak scale) 
the new layer of physical reality might lie much above the TeV scale. 
This is actually suggested by the inner structure of the SM: 
flavour and CP violating observables have generically 
probed scales up to $\Lambda_{\rm SM} \gtrsim 10^{6}$ GeV, 
while light neutrino masses point to $\Lambda_{\rm SM} \lesssim 10^{14}$ GeV. 
At the same time, 
it is evident that the hypercharge structure of the 
SM fermions cries out for unification 
(letting aside the 
more mysterious 
origin of flavour). 
Left-Right symmetric theories \cite{Pati:1974yy,Mohapatra:1974gc,Senjanovic:1975rk,Mohapatra:1979ia} 
provide a most natural route 
for addressing the origin of hypercharge and neutrino masses, 
passing through the Pati-Salam partial unification scheme \cite{Pati:1974yy} 
(which also provides a rationale for the quantization of electric charge)
and ending up into one SM family plus a right-handed neutrino unified into 
a spinorial representation of SO(10) \cite{Fritzsch:1974nn,Georgi:1974my}. 
Due to the fact that these groups have rank 5, they admit at least an intermediate 
breaking stage before landing on the SM gauge group, and in the case of 
Pati-Salam \cite{Melfo:2003xi} and SO(10) \cite{Chang:1984qr,Deshpande:1992em,Bertolini:2009qj} 
those are often predicted to lie in between 
$10^{6}$ GeV and $10^{14}$ GeV by (partial) gauge coupling unification. 
This picture would gain an additional value if such intermediate-scale physics would be 
connected to other open issues of the SM, most notably the baryon asymmetry 
of the Universe and Dark Matter (DM). 
The former is built-in in the form of thermal leptogenesis \cite{Fukugita:1986hr}, 
which in its simplest realization would suggest $\Lambda_{\rm SM} \gtrsim 10^9$ GeV 
(see e.g.~\cite{Davidson:2008bu}), 
while DM is often a missing ingredient in minimal 
realizations of Left-Right symmetric theories 
(for an exception, see \cite{Nemevsek:2012cd}). 
A natural possibility is then to impose a Peccei-Quinn (PQ) symmetry \cite{Peccei:1977hh,Peccei:1977ur}
delivering an axion \cite{Weinberg:1977ma,Wilczek:1977pj}, 
which provides at the same time an excellent DM candidate \cite{Preskill:1982cy,Abbott:1982af,Dine:1982ah}
and solves the strong CP problem. 
This choice is economical also in the following sense: 
$i)$ in SO(10) setups the PQ symmetry was often imposed for another reason, 
namely to enhance the predictivity of 
the renormalizable (non-supersymmetric) 
SO(10) Yukawa sector \cite{Babu:1992ia} and 
$ii)$ it is based on a coincidence of scales: the axion decay constant is in fact 
bounded from astrophysical and cosmological consideration within the range 
$10^8 \ \text{GeV} \lesssim f_a \lesssim 10^{18} \ \text{GeV}$ 
(see \cite{DiLuzio:2020wdo} for updated limits). 

The scenario above is both 
simple and elegant enough 
for it to be taken seriously. 
Given the fact that it was clearly envisioned around four decades ago 
by Mohapatra and Senjanovi\'c \cite{Mohapatra:1982tc} 
in the context of SO(10), 
and often reconsidered in different ways 
and at different levels of depth \cite{Davidson:1981zd,Reiss:1981nd,Lazarides:1981kz,Holman:1982tb,Kalara:1983ye,Davidson:1983fy,Davidson:1983fe,Chang:1987hz,Fukuyama:2004zb,Bajc:2005zf,Altarelli:2013aqa,Babu:2015bna,Saad:2017pqj,Ernst:2018bib,Ernst:2018rod,Boucenna:2018wjc,Babu:2018qca,Hamada:2020isl,Lazarides:2020frf}, it is worth to spend few words on why again now and why 
the Pati-Salam axion. 

From an experimental point of view, there are now better hopes to catch the axion tail of 
the story. Axion physics is in fact in a blooming phase 
with several new detection concepts which promise to open for explorations 
regions of parameter space which were thought unreachable until few years ago
(for updated experimental reviews see \cite{Sikivie:2020zpn,Irastorza:2018dyq}). 
In particular, the possibility that the Grand Unified Theory (GUT) axion window 
could be completely covered by the axion DM experiments 
CASPEr-Electric \cite{Budker:2013hfa,JacksonKimball:2017elr}
and ABRACADABRA \cite{Kahn:2016aff} has triggered a revival of 
studies of GUT$\times \U(1)_{\rm PQ}$ models, 
with the axion field residing in a non-singlet representation of the GUT group. 
In particular, Ref.~\cite{Ernst:2018bib} computed for the first time 
low-energy axion couplings in SO(10)$\times \U(1)_{\rm PQ}$ models 
and considered axion mass predictions in SO(10) models with up to two intermediate breaking stages. 
Ref.~\cite{DiLuzio:2018gqe} considered instead a minimal non-renormalizable 
SU(5)$\times \U(1)_{\rm PQ}$ model based on
a PQ extension of \cite{Bajc:2006ia,Bajc:2007zf}, 
which due to its minimality allowed 
to obtain 
(via the three-loop gauge coupling unification analysis of \cite{DiLuzio:2013dda})
a sharp prediction for the axion mass 
in the neV domain. Subsequently, Refs.~\cite{FileviezPerez:2019fku,FileviezPerez:2019ssf} 
considered axion mass predictions in other minimal 
renormalizable SU(5)$\times \U(1)_{\rm PQ}$ models.  
Some cosmological consequences of 
supersymmetric axion GUTs 
were considered instead in \cite{Co:2016xti}.    

The study of the Pati-Salam axion considered in the present work has a twofold motivation. 
On the one hand, the Pati-Salam (partial) unification constraints are genuinely different from 
SO(10) ones, so their predictions can be in principle discerned from those of SO(10). 
For instance, while it is notoriously difficult to obtain a low-scale Left-Right symmetry breaking scale 
in SO(10), we will show that if the Pati-Salam group is broken at the Planck scale, 
the Left-Right symmetry breaking can be as low as 20 TeV (and even lower in the absence of the PQ).  
One the other hand, the Pati-Salam gauge group, 
which is half-way through SO(10), 
provides a simpler setup 
and for this reason it can be studied in quite some detail. 
For instance, although SO(10)$\times \U(1)_{\rm PQ}$ scalar potentials 
have been partially classified in \cite{Lazarides:1981kz}, 
they have never been investigated in detail. 
In the present work we provide also a non-trivial step in that direction, 
by working out the scalar potential dynamics of a complex scalar adjoint of $\SU(4)_{\rm PS}$, 
which hosts the axion field as a phase, 
and whose vacuum expectation value (VEV) 
is simultaneously responsible for PQ and Pati-Salam breaking down to the Left-Right symmetric gauge group. This allows in turn to constrain the axion mass via a 
renormalization group (RG) analysis of (partial) gauge coupling unification in Pati-Salam. 

The paper is structured as follows. 
In \sect{sec:minimalmodelconstr} we describe 
the logic behind the construction of a minimal renormalizable 
Pati-Salam$\times \U(1)_{\rm PQ}$ model.  
Next, we focus on axion couplings (\sect{sec:axioncoupl}) and 
on the axion mass prediction from Pati-Salam breaking (\sect{sec:PSbreakdyn}). 
This is the main conceptual point of the paper, which 
relies on the calculable relation between the axion mass  
and the Pati-Salam breaking scale (cf.~\eq{eq:mavsMPS}). 
The latter can then be constrained via a RG analysis 
of (partial) gauge coupling unification within the Pati-Salam model. 
In \sect{sec:RGE} we report the results of such RG analysis, 
which is based both on a one-loop analytical understanding and a more 
involved two-loop numerical investigation, whose details are deferred to \app{app:2loopbf}. 
The main outcome of the RG analysis is that the 
Pati-Salam breaking scale (the axion mass) becomes a decreasing (increasing) 
function of a sliding Left-Right symmetry breaking scale. 
\sect{sec:pheno} is devoted to the phenomenology of the model and to the 
experimental prospects for hunting the Pati-Salam axion. 
We first collect various cosmological and astrophysical constraints 
(\sect{sec:astrocosmo})
and then review (\sect{eq:axionDM}) 
the future sensitivity of the axion DM experiments 
ABRACADABRA and 
CASPEr-Electric.
The main outcome is that the parameter space of the Pati-Salam axion 
will be fully covered by the late phases of those axion DM experiments, 
as shown in \fig{fig:maxionvsMLR}.  
We finally consider possible 
correlated signals due to Pati-Salam (\sect{eq:PSLRsignatures}) 
and Left-Right (\sect{eq:LRonlysignatures}) 
symmetry breaking dynamics. 
While the former are more difficult to be observable, 
a sliding Left-Right breaking scale can give rise to more interesting indirect/direct signatures. 
We conclude in \sect{sec:concl} with a brief recap of the main results, 
together with a discussion of the critical points of the present setup and 
an outlook for possible future work.


\section{Peccei-Quinn extended Pati-Salam} 
\label{sec:PQextPS}

In this Section we propose a simple 
implementation of the PQ symmetry in a minimal Pati-Salam model, 
which is inspired by the more studied case of $\SO(10) \times \U(1)_{\rm PQ}$ 
\cite{Mohapatra:1982tc,Bajc:2005zf,Altarelli:2013aqa,Ernst:2018bib}.  
A similar Pati-Salam$\times \U(1)_{\rm PQ}$ construction 
has been recently considered in Ref.~\cite{Saad:2017pqj}. 
The main difference compared to \cite{Saad:2017pqj} 
is that here the axion field resides in a 
non-singlet representation of Pati-Salam. 
This allows in turn to connect the axion mass 
and the Pati-Salam breaking scale, 
with the latter being constrained 
via a RG analysis of gauge coupling 
(partial) unification in Pati-Salam.

\subsection{Minimal model construction}
\label{sec:minimalmodelconstr}

The Pati-Salam gauge group is defined by 
\beq 
\label{eq:defGPS}
\G_{\rm PS} \equiv \SU(4)_{\rm PS} \times \SU(2)_L \times \SU(2)_R \times \P \, , 
\eeq
where $\P$ is a discrete symmetry exchanging $L \leftrightarrow R$, 
which enforces parity restoration in the UV. 
The color factor is embedded as $\SU(3)_C \times \U(1)_{B-L} \subset \SU(4)_{\rm PS}$, thus implementing the 
idea of lepton number as the fourth color \cite{Pati:1974yy}.\footnote{Although the original formulation 
was based on the gauge group $\SU(4)_{\rm PS} \times \SU(4)_L \times \SU(4)_R$, 
the 
$\SU(4)_{\rm PS} \times \SU(2)_L \times \SU(2)_R$ setup emerged 
later on as a simpler UV completion of the SM.} 
The embedding of the hypercharge follows the standard one of Left-Right 
symmetric theories \cite{Pati:1974yy,Mohapatra:1974gc,Senjanovic:1975rk,Mohapatra:1979ia}  
\beq 
\label{eq:YrelationLR}
Y = T^3_R + \frac{B-L}{2} \, . 
\eeq
The fermion fields transform under $\G_{\rm PS}$ 
as $\Q_L \sim (4,2,1)$ and $\Q_R \sim (4,1,2)$. Explicitly, 
the embedding of the SM fermions consists in three copies of
\beq 
\label{eq:SMfermembed}
\Q_L = 
\begin{pmatrix}
u^1_L & u^2_L & u^3_L & \nu_L \\ 
d^1_L & d^2_L & d^3_L & e_L 
\end{pmatrix} \, , \qquad 
\Q_R = 
\begin{pmatrix}
u^1_R & u^2_R & u^3_R & \nu_R \\ 
d^1_R & d^2_R & d^3_R & e_R 
\end{pmatrix} \, , 
\eeq 
including also a RH neutrino, $\nu_R$. 
Under $\P: \Q_L \leftrightarrow \Q_R$, so that $\P$ assumes the meaning of space-time 
parity. Instead of $\P$ one could consider charge conjugation, 
$\C: \Q_L \leftrightarrow \Q^c_R$. 
This latter choice is ultraviolet (UV) 
motivated by the fact that, 
as originally observed in \cite{Kibble:1982ae,Kibble:1982dd}, 
$\C$ turns out to be an element of SO(10) 
(see also \cite{Chang:1983fu,Chang:1984uy}). 
The difference between the two choices mainly regards the structure of CP violation 
(see e.g.~\cite{Maiezza:2014ala,Bertolini:2019out,Senjanovic:2020int}).\footnote{In particular, as pointed out in Ref.~\cite{Maiezza:2014ala}, 
imposing a PQ symmetry 
in the case of $\P$ allows to non-trivially relax the strong bounds 
from $\epsilon_K$ on the Left-Right symmetry breaking scale.} 
Since the key results of this work concerning Pati-Salam breaking dynamics
will not be affected by this choice, we will stick for definiteness to the case of $\P$.

The Higgs sector comprises the following representations: 
$\S \sim (15,1,1)$,
$\Delta_L \sim (10,3,1)$,  
$\Delta_R \sim (10,1,3)$, 
$\Phi_{1} \sim (1,2,2)$, $\Phi_{15} \sim (15,2,2)$. 
Let us motivate in turn the need for such representations: 
$\S$ is introduced in order to allow for an intermediate breaking stage: 
\beq 
\label{eq:intermediateLR}
\G_{\rm PS} \xrightarrow[]{\vev{\S}} \G_{\rm LR} \equiv \SU(3)_{C} \times \SU(2)_L \times \SU(2)_R \times \U(1)_{B-L} \times \P \, . 
\eeq 
We will also consider the case in which the VEV of $\S$ breaks spontaneously the symmetry $\P$, 
so that the unbroken group is $\G^{\slashed{P}}_{\rm LR} \equiv \SU(3)_{C} \times \SU(2)_L \times \SU(2)_R \times \U(1)_{B-L}$. The field $\Delta_R$ is needed to provide the final symmetry breaking stage down to the 
SM gauge group
\beq 
\label{eq:intermediateLR}
\G_{\rm LR}^{(\slashed\P)} \xrightarrow[]{\vev{\Delta_R}} \G_{\rm SM} \equiv \SU(3)_{C} \times \SU(2)_L \times \U(1)_Y \, . 
\eeq
The field $\Delta_L$ is required 
in order to make the theory Left-Right symmetric under $\P:\Delta_L \leftrightarrow \Delta_R$. 
Finally, the (complex) 
bi-doublets $\Phi_1$ and $\Phi_{15}$ are needed in order to reproduce SM fermion masses 
and mixing. The renormalizable Yukawa Lagrangian reads\footnote{In order to ease 
the notation, flavour, gauge and Lorentz contractions are left understood.}
\begin{align}
\label{eq:LYPS}
\mathcal{L}_Y &= \bar \Q_L ( Y_{1} \Phi_1 + Y_{15} \Phi_{15} + \tilde Y_{1} \tilde \Phi_1 + \tilde Y_{15}\tilde  \Phi_{15} ) \Q_R \nonumber \\
&+ Y_{\Delta_L} \Q_L \Q_L \Delta_L + Y_{\Delta_R} \Q_R \Q_R \Delta_R + \text{h.c.} \, ,
\end{align}
where $\tilde \Phi_{1,15} = \epsilon\, \Phi^*_{1,15}  \epsilon$ (with $\epsilon = i \sigma_2$) 
denote conjugate bi-doublet fields. 
Further assuming $\P: \Phi_{1,15} \to \Phi^\dag_{1,15}$, the invariance of 
$\mathcal{L}_Y$ under $\P$ requires $Y_{1,15} = Y^\dag_{1,15}$, $\tilde Y_{1,15} = \tilde Y^\dag_{1,15}$, 
$Y_{\Delta_L} = Y_{\Delta_R} \equiv Y_{\Delta}$ (with $Y_{\Delta} = Y_{\Delta}^T$). 

In fact, without $\Phi_{15}$ one would end up with 
a wrong mass relation between down-quarks and charged leptons, $M_d = M_e^T$. 
This is avoided 
by introducing $\Phi_{15}$ which transforms non-trivially under $\SU(4)_{\rm PS}$, 
so that after $\SU(4)_{\rm PS}$ breaking $\vev{\Phi_{15}}$ feeds differently into down-quarks and charged leptons \cite{Pati:1983zp}. 

On the other hand, the proliferation of Yukawa matrices in \eq{eq:LYPS} makes 
the model not predictive for fermions masses and mixings. 
This provides a rationale for introducing (similarly as was 
originally proposed for the SO(10) Yukawa sector \cite{Babu:1992ia})
a $\U(1)_{\rm PQ}$: 
\begin{align} 
\label{eq:PQcharges}
\Q_L &\to e^{i\frac{\alpha}{2}} \Q_L \, ,  \quad 
\Q_R \to e^{-i\frac{\alpha}{2}} \Q_R \, ,  \quad 
\Phi_{1} \to e^{i\alpha} \Phi_{1} \, , \quad 
\Phi_{15} \to e^{i\alpha} \Phi_{15} \, ,  \nonumber \\
\Delta_L &\to e^{-i\alpha} \Delta_L \, ,  \quad 
\Delta_R \to e^{i\alpha} \Delta_R \, ,  
\end{align}
which enhances the predictivity of the Pati-Salam Yukawa sector by enforcing
$\tilde Y_{1,15} \to 0$. After electroweak symmetry breaking, one obtains the following SM fermion mass 
sum rules (respectively for the up-, down-quarks, charged-leptons, 
Dirac neutrinos, left-handed and right-handed Majorana neutrinos)\footnote{The $-3$ factor for the leptonic components in \eqs{eq:SMSRMe}{eq:SMSRMD} 
can be understood from the fact that $\vev{15} \propto \diag(1,1,1,-3)$ in $\SU(4)_{\rm PS}$ space.}
\begin{align}
\label{eq:SMSRMu}
M_u &= Y_1 v_1^u + Y_{15} v_{15}^u \, , \\
\label{eq:SMSRMd}
M_d &= Y_1 v_1^d + Y_{15} v_{15}^d \, , \\
\label{eq:SMSRMe}
M_e &= Y_1 v_1^d -3 Y_{15} v_{15}^d \, , \\
\label{eq:SMSRMD}
M_D &= Y_1 v_1^u -3 Y_{15} v_{15}^u \, , \\
\label{eq:SMSRML}
M_L &= Y_\Delta v_L \, , \\
\label{eq:SMSRMR}
M_R &= Y_\Delta v_R \, , 
\end{align}
where we introduced the VEVs: 
\beq 
\vev{\Phi_{1,15}} = 
\begin{pmatrix}
v^u_{1,15} & 0 \\
0 & v^d_{1,15}
\end{pmatrix} \, , \qquad 
\vev{\Delta_{L,R}} = 
\begin{pmatrix}
0 & 0 \\
v_{L,R} & 0
\end{pmatrix} \, ,
\eeq
with $(v_1^u)^2 + (v_{15}^u)^2 + (v_1^d)^2 + (v_{15}^d)^2 \equiv v^2 = (174 \ \text{GeV})^2$. 
Light neutrino mass eigenstates follow the standard type-I+II seesaw formula 
\beq 
\label{eq:SMSRnu}
M_\nu = M_L - M_D M_R^{-1} M_D^T \, . 
\eeq
From the above mass relations we qualitatively conclude that since $M_u$ and $M_D$ are 
strongly correlated, the top mass eigenvalue prefers an intermediate scale $v_R \gg v$, 
unless a tuning is invoked into the Dirac neutrino mass term, 
$M_D$.\footnote{Ref.~\cite{Saad:2017pqj} performed the 
fit to the Pati-Salam fermion mass sum rules fixing $v_R = 10^{14}$ GeV and 
assuming the dominance of Type-I seesaw. Similar fermion mass sum rules in the (more constraining) 
case of SO(10) are also known to yield viable fits (see e.g.~\cite{Joshipura:2011nn}). 
A quantitative assessment of fermion masses and mixing in the minimal Pati-Salam model, 
leaving the $v_R$ scale free, is left for a future study.} 
Moreover, 
$v_L \sim v^2 / v_R$ is an induced VEV from the minimization of the scalar potential, 
which is consistent with the requirement of light neutrino masses. 
We hence assume the hierarchy of scales $v_\S > v_R \gg v \gg v_L$ (where we also introduced 
the $\G_{\rm PS} \times \U(1)_{\rm PQ} \to \G^{(\slashed{P})}_{\rm LR}$ VEV 
$\vev{\S} = v_\S$).  

In the present setup, where only the fields in \eq{eq:PQcharges} are charged under PQ, 
the VEV $\vev{\Delta_R}$ breaks $\SU(2)_R \times \U(1)_{B-L} \times \U(1)_{\rm PQ}$ down to 
$\U(1)_Y \times \U(1)'_{\rm PQ}$, where $\U(1)'_{\rm PQ}$ is a new global PQ symmetry 
(a linear combination of the original PQ and the broken gauge generators), 
which is eventually broken at the electroweak scale by $\vev{\Phi_{1,15}}$, 
thus leading to an experimentally untenable Weinberg-Wilczek axion.  
A natural way to fix this is to complexify the representation $\S$ and charge it under the 
$\U(1)_{\rm PQ}$, thus connecting the $\G_{\rm PS} \to \G^{(\slashed{P})}_{\rm LR}$ breaking scale 
with the PQ breaking scale. A possible choice is 
\beq 
\label{eq:PQSfield}
\S \to e^{ i\alpha} \S \, ,
\eeq
and the scalar potential can be written as 
\beq 
\label{eq:scalpot}
\V = \V_{\rm r.i.} + \V_{\rm PQ} \, , 
\eeq
where $\V_{\rm r.i.}$ contains re-phasing invariant terms which are not sensitive to $\U(1)$ phases, 
while $\V_{\rm PQ}$ is chosen in such a way to ensure the explicit breaking
\beq 
\label{eq:explicitU(1)break}
\U(1)_{\S} \times \U(1)_{\Delta_L}  \times \U(1)_{\Delta_R} \times \U(1)_{\Phi_1}  \times \U(1)_{\Phi_{15}} 
\to \U(1)_{B-L} \times \U(1)_{\rm PQ} \, ,
\eeq
and reads (see also \cite{Saad:2017pqj})
\begin{align}
\label{eq:VPQ}
\V_{\rm PQ} &= 
\lambda_{1} \S^{\dag 2} \Phi_1^2 
+ \lambda_{15} \S^{\dag 2} \Phi_{15}^2 
+  \lambda_{\rm mix} \, \S^{\dag2} \Phi_{1} \Phi_{15} 
+ \alpha_{\rm mix} \Phi^{\dag 2}_1 \Phi^2_{15}
\nonumber \\
&+ \beta_{1}  \Phi_1^\dag \Delta_R \Phi_1^\dag \Delta_L^\dag 
+ \beta_{15}  \Phi_{15}^\dag \Delta_R \Phi_{15}^\dag \Delta_L^\dag
+ \beta_{\rm mix}  \Phi_{1}^\dag \Delta_R \Phi_{15}^\dag \Delta_L^\dag
\nonumber \\
&
+ \gamma_{15} ( \Phi_{15}^{\dag 2} \Delta_R^2 +  \Phi_{15}^2 \Delta_L^2 )
+ \gamma_{\rm mix} ( \Phi_{1}^\dag \Phi_{15} \Delta_R \Delta_R^\dag 
+  \Phi_{1}  \Phi_{15}^\dag \Delta_L \Delta_L^\dag) \nonumber \\
&
+ \delta_{15}  \Phi_{15}^\dag \Phi_{15} \Delta_R \Delta_L 
+ \eta \Delta_R^2 \Delta_L^2
+ \omega ( \S^{\dag 2} \Delta_R^2 +  \S^2 \Delta_L^2 ) 
+ \text{h.c.} 
\, , 
\end{align}
where multiple gauge invariant contractions with the same $\U(1)_{\rm PQ}$ structure 
are left understood.\footnote{For instance, $\S^{*2} \Phi_{15}^2$ features four linearly-independent 
invariants: 
$[\S^{*2}]_1 [\Phi_{15}]_1^2$, 
$[\S^{*2}]_{15} [\Phi_{15}]_{15}^2$,
$[\S^{*2}]_{20'} [\Phi_{15}]_{20'}^2$ 
and $[\S^{*2}]_{84} [\Phi_{15}]_{84}^2$  
where subscripts denote the type of $\SU(4)_{\rm PS}$ 
contraction. 
Since in this work we will not address the full minimization of the 
$\G_{\rm PS} \times \U(1)_{\rm PQ}$ 
scalar potential, such details are not essential for the following discussion.} 
The $\lambda_{1,15}$ terms are needed to communicate the PQ breaking from the $\S$ to the 
bi-doublets $\Phi_{1,15}$, $\lambda_{\rm mix}$ and $\alpha_{\rm mix}$
are allowed by the global symmetries left invariant 
by $\lambda_{1,15}$, while $\beta_{1,15,\,\text{mix}} \neq 0$ is required in order to avoid 
an extra spontaneously broken U(1) global symmetry, with an associated (unwanted) Goldstone boson. 
Other terms in the third and fourth line of \eq{eq:VPQ} are allowed by 
gauge invariance and the global symmetries of the system. 
$\P$ invariance (with $\P: \S \to \S^\dag$ being a possible definition 
compatible with the PQ symmetry) implies 
some restrictions on the scalar potential parameters,  
e.g., the couplings in the first two rows of \eq{eq:VPQ} need to be real 
(since the h.c.~operator coincides with the $\P$-transformed one).
The field transformation properties under $\G_{\rm PS} \times \U(1)_{\rm PQ}$ are collected for convenience 
in \Table{tab:PSfields}.
\begin{table}[htp]
\begin{center}
\begin{tabular}{|c|c|c|c|c|c|}
\hline
Field & $\SU(4)_{\rm PS}$ & $\SU(2)_L$ & $\SU(2)_R$ & $\P$ & $\U(1)_{\rm PQ}$ \\
\hline
$\Q_L$ & $4$ & $2$ & $1$ & $\Q_L \to \Q_R$ & $\frac{1}{2}$ \\
$\Q_R$ & $4$ & $1$ & $2$ & $\Q_R \to \Q_L$ & $-\frac{1}{2}$ \\
\hline
$\Phi_1$ & $1$ & $2$ & $2$ & $\Phi_1 \to \Phi^\dag_1$ & $1$ \\ 
$\Phi_{15}$ & $15$ & $2$ & $2$ & $\Phi_{15} \to \Phi^\dag_{15}$ & $1$ \\ 
$\Delta_L$ & $10$ & $3$ & $1$ & $\Delta_L \to \Delta_R$ & $-1$ \\ 
$\Delta_R$ & $10$ & $1$ & $3$ & $\Delta_R \to \Delta_L$ & $1$ \\ 
$\S$ & $15$ & $1$ & $1$ & $\S \to \S^\dag$ & $1$ \\
\hline
\end{tabular}
\end{center}
\caption{Field content of the minimal $\G_{\rm PS} \times \U(1)_{\rm PQ}$ model.}
\label{tab:PSfields}
\end{table}


\subsection{Axion couplings}
\label{sec:axioncoupl}

In the presence of spontaneously 
broken gauge symmetries the identification of the canonical axion field and its couplings to SM fields 
presents some non-trivial steps. The axion field must be properly orthogonalized in order to avoid kinetic 
mixings with the would-be Goldstone bosons associated with broken Cartan generators \cite{Ernst:2018bib}. 
Due to the hypercharge relation in Left-Right symmetric models (cf.~\eq{eq:YrelationLR}), 
it is indeed enough to require orthogonality with respect to $T^3_L$ and $T^3_R$ \cite{BDLN}. 
This is particularly relevant for the axion coupling to matter fields (nucleons and electrons)
which become functions of vacuum angles, expressed in terms of the gauge symmetry breaking VEVs. 
Since these couplings will not be phenomenologically relevant for axion mass range 
discussed in this paper, we will not report here their derivation, but just quote the final results 
(for a more detailed account, see \cite{BDLN}).  
On the other hand, the axion coupling to photons depends only on the anomaly coefficients 
of the PQ current, defined via 
\beq 
\label{eq:dJPQAnomal}
\partial^\mu J^{\rm PQ}_{\mu} = 
\frac{\alpha_s N}{4\pi} 
G \tilde G
+\frac{\alpha E}{4\pi} 
F \tilde F
\, , 
\eeq
which can be actually computed in terms of the $\U(1)_{\rm PQ}$ charges of 
\Table{tab:PSfields} (generically denoted as $\X_i$), and are found to be 
(see e.g.~\cite{Srednicki:1985xd})
\begin{align}
\label{eq:N}
N &= n_g \times 2 T(3) \times ( \X_{\Q_L} - \X_{\Q_R} ) = 3 \\ 
\label{eq:E}
E &= n_g \times (3 Q_u^2 + 3 Q_d^2 + Q_e^2) \times (\X_{\Q_L} - \X_{\Q_R}) = 8
\end{align}
where we used $n_g = 3$ (number of generations), 
$T(3) = \tfrac{1}{2}$ (Dynkin index of the fundamental of $\SU(3)_C$) 
and the electric charges $Q_u = \tfrac{2}{3}$, $Q_d = -\tfrac{1}{3}$,  $Q_e = -1$.   
Hence, in particular $E/N = 8/3$, which sets the axion coupling to photons (see below).  

The axion effective Lagrangian, including couplings to photons, matter fields ($f=p,n,e$) 
and the oscillating neutron Electric Dipole Moment (nEDM), 
can be written as
\begin{equation} 
\label{eq:Laint1}
\mathcal{L}^{\rm int}_a = \frac{\alpha}{8 \pi} \frac{C_{a\gamma}}{f_a} a F \tilde F
+ C_{af} \frac{\partial_\mu a}{2 f_a} \bar f \gamma^\mu \gamma_5 f 
- \frac{i}{2} \frac{C_{an\gamma}}{m_n} \frac{a}{f_a}  \bar n \sigma_{\mu\nu} \gamma_5 n F^{\mu\nu} 
\, ,   
\end{equation}
with the values of the $C_{ax}$ coefficients given by \cite{diCortona:2015ldu,Pospelov:1999ha,DiLuzio:2020wdo}
\begin{align}
\label{eq:Cagamma}
C_{a\gamma} &= \frac{E}{N} - 1.92(4) \, , \\
\label{eq:Cap}
C_{ap} &= -0.47(3) + 0.88(3) \, c^0_u - 0.39(2) \, c^0_d - C_{a,\, \text{sea}}
\, , \\
\label{eq:Can}
C_{an} &= -0.02(3) + 0.88(3) \, c^0_d - 0.39(2) \, c^0_u - C_{a,\, \text{sea}}
\, , \\ 
\label{eq:Casea}
C_{a,\, \text{sea}} &= 0.038(5) \, c^0_s 
+0.012(5) \, c^0_c + 0.009(2) \, c^0_b + 0.0035(4) \, c^0_t \, , \\
\label{eq:Cae}
C_{ae} &= c_e^0 + \frac{3\alpha^2}{4\pi^2} 
\[ \frac{E}{N} \log\( \frac{f_a}{m_e} \)
- 1.92(4)
\log\( \frac{\text{GeV}}{m_e} \) \] \, , \\
\label{eq:Cangamma}
C_{an\gamma} &= 0.011(5) \, e \, , 
\end{align} 
in terms of model-dependent factors $E/N$ and $c^0_{u,\,d,\,e}$. 
In the Pati-Salam axion model they are found to be (up to safely negligible 
$v^2 / v^2_{\S} \ll 1$ corrections for the fermion couplings, see \cite{BDLN})
\beq
\label{eq:PSaxioncoupl}
E/N = 8/3 \, , \quad 
c^0_{u_i} = \frac{1}{3} \sin^2\beta \, , \quad
c^0_{d_i} = \frac{1}{3} \cos^2\beta \, , \quad  
c^0_{e_i} = \frac{1}{3} \cos^2\beta \, , \\
\eeq
with the index $i=1,2,3$ denoting generations and 
\beq 
\tan{\beta} = \sqrt{\frac{(v_1^d)^2 + (v_{15}^d)^2}{(v_1^u)^2 + (v_{15}^u)^2}} \, .
\eeq

\subsection{Pati-Salam breaking dynamics}
\label{sec:PSbreakdyn}

The complex $\SU(4)_{\rm PS}$ adjoint representation $\S$ is responsible for the 
initial breaking 
\beq 
\label{eq:SU4break}
\SU(4)_{\rm PS} \times \U(1)_{\rm PQ} 
\xrightarrow[]{\vev{\S}} 
\SU(3)_{C} \times \U(1)_{B-L} \, .
\eeq 
Since $\vev{\S}$ provides the largest VEV, this dynamics is captured by the 
$\S$ sector of the re-phasing invariant potential (here we restore gauge 
contractions in $\SU(4)_{\rm PS}$ space) 
\begin{align} 
\label{eq:VmodS}
\V_{\rm r.i.} &\supset 
-\mu^2_\S \Tr \S^\dag \S 
+ \lambda^{(1)}_{\S}  (\Tr \S^\dag \S)^2
+ \lambda^{(2)}_{\S}  (\Tr \S^\dag \S^\dag) (\Tr \S \S) \nonumber \\ 
&+ \lambda^{(3)}_{\S}  \Tr \S^\dag \S \S^\dag \S 
+ \lambda^{(4)}_{\S}  \Tr \S^\dag \S^\dag \S \S 
+ \lambda^{(5)}_{\S} [\epsilon \epsilon \S \S^\dag \S \S^\dag] 
\, ,
\end{align} 
where the last invariant reads explicitly 
\beq 
[\epsilon \epsilon \S \S^\dag \S \S^\dag] \equiv 
\epsilon_{ijkl} \epsilon^{mnop} (\S)^i_m (\S^\dag)^j_n (\S)^k_o (\S^\dag)^l_p \, ,
\eeq
and we have set to zero operators of the type $\Tr \S^n$, 
with $n=2,3,4$, consistently with the presence of the PQ symmetry (cf.~\eq{eq:PQSfield}). 
We decompose the complex adjoint in terms of canonically normalized real fields as 
(see also \cite{Buttazzo:2019mvl})
\beq 
\label{eq:Sdec}
\S = \[ (v_{\S} + \rho) T^{15} + \sum_{b=1}^{14} (\phi^b_R + i\phi^b_I) T^{b} \]  e^{i \frac{a}{v_\S}} \, , 
\eeq 
with $\SU(4)_{\rm PS}$ generators (see e.g.~Appendix A.10 of \cite{DiLuzio:2018zxy}), 
normalized as $\Tr T^\alpha T^\beta = \frac{1}{2} \delta^{\alpha\beta}$. 
We assume $\vev{\S} = T^{15} v_{\S}$, with 
\beq 
T^{15} = \tfrac{1}{2\sqrt{6}} \diag(1,1,1,-3) \, ,
\eeq
then the unbroken generators acting trivially on the vacuum,  
$[T^{\alpha}, \vev{\S}] = 0$, span an $\SU(3)_C$ algebra ($\alpha=1,\ldots,8$) times  
$T^{15}$, which is identified (up to an overall normalization) with $\U(1)_{B-L}$. 
Schematically, the decomposition of a complex $\SU(4)_{\rm PS}$ adjoint 
under $\SU(3)_{C} \times \U(1)_{B-L}$ reads
\beq 
\label{eq:Sdecschem}
15 \to (8,0) + (1,0) + (3,\tfrac{4}{3}) + (\bar 3,-\tfrac{4}{3}) \, . 
\eeq

\subsubsection{Scalar boson spectrum}

The calculation of the scalar spectrum for $\mu^2_\S > 0$ yields:
\begin{itemize}
\item A (perturbatively) massless axion: $m_a^2 = 0$;
\item A radial mode: $m_\rho^2 = 
\tfrac{1}{6}
( 12 ( \lambda^{(1)}_{\S} + \lambda^{(2)}_{\S} )
+ 7 (\lambda^{(3)}_{\S} 
+ \lambda^{(4)}_{\S}) 
- 6 \lambda^{(5)}_{\S}
) v^2_\S$;
\item 6 would-be Goldstone modes (eaten by the massive vector leptoquark $\X_\mu \sim (3,\tfrac{4}{3})$ 
(under $\SU(3)_{C} \times \U(1)_{B-L}$)): 
$m^2_{\phi^A_R} = 0$ (for $A=9,\ldots,14$); 
\item Three sets of degenerate scalar modes for a total of 22 massive scalars, with masses:
\begin{itemize}
\item $m^2_{\phi^a_R} = -\frac{1}{3} (\lambda^{(3)}_{\S} + \lambda^{(4)}_{\S} 
- 6 \lambda^{(5)}_{\S}
)  v^2_\S$ (for $a=1,\ldots,8$), 
\item $m^2_{\phi^a_I} = -\frac{1}{2} (4\lambda^{(2)}_{\S} + \lambda^{(3)}_{\S} + \lambda^{(4)}_{\S} 
- 2 \lambda^{(5)}_{\S})  v^2_\S$ 
(for $a=1,\ldots,8$), 
\item $m^2_{\phi^A_I} = - \frac{1}{6}  (12 \lambda^{(2)}_{\S} 
- 3 \lambda^{(3)}_{\S} 
+ 5 \lambda^{(4)}_{\S}  
- 2 \lambda^{(5)}_{\S})  v^2_\S$ 
(for $A=9,\ldots,14$).
\end{itemize}
\end{itemize}
The conditions on the scalar potential parameters leading to a 
positive mass spectrum are straightforward and they serve 
to show that the $\SU(4)_{\rm PS} \times \U(1)_{\rm PQ} \to \SU(3)_{C} \times \U(1)_{B-L}$ configuration can be at least a local minimum. 
Since they are lengthy and not needed in the following, we do not 
report them explicitly. 


\subsubsection{Gauge boson spectrum}
The gauge boson spectrum can be determined from the action of the covariant derivative, 
$D_\mu \Phi = \partial_\mu + i g_{\rm PS} [T^\alpha, \Phi] \A^\alpha_\mu$, and the kinetic term 
\beq
\Tr (D_\mu \S)^\dag (D^\mu \S) \supset \frac{1}{2} 
\( 2 g^2 \Tr [T^\alpha, \vev{\S}]^\dag [T^\beta, \vev{\S}] \) 
\A^\alpha_\mu \A^{\beta\mu} \, .  
\eeq
We find: 
\begin{itemize}
\item A vector leptoquark $\X_\mu  \sim (3,\tfrac{4}{3})$ spanning over $\A_\mu^A$ (for $A=10,\ldots,14$): 
$m^2_{\X} = \tfrac{2}{3} g_{\rm PS}^2 v_{\S}^2$; 
\item $8+1$ massless gauge boson associated with the unbroken $\SU(3)_C \times \U(1)_{B-L}$ 
algebra.
\end{itemize}

\subsubsection{Axion mass from Pati-Salam breaking scale}

The axion field resides dominantly in $\S$, hence (up to $(v_R/v_\S)^2$ and $(v/v_\S)^2$ 
corrections\footnote{For $v_R \approx v_\S$ (one-step breaking) 
$\G_{\rm LR}$-breaking dynamics should be taken as well into account  
for the proper matching of the axion decay constant with the symmetry breaking VEVs.})
\beq 
f_a \approx \frac{v_{\S}}{2N} \, , 
\eeq
where $2N = 6$ (see \eq{eq:N}).
At the same time the Pati-Salam partial-unification scale, $M_{\rm PS}$, can be identified with the mass of the vector leptoquark\footnote{This identification is valid up to scalar threshold effects 
(cf.~\eqs{eq:1loopmatchgen}{eq:1loopmatchgen2}).} 
\beq 
\label{eq:MPSfa}
M_{\rm PS} \approx m_{\X} = \sqrt{\frac{2}{3}} g_{\rm PS} v_\S 
\approx 2 \sqrt{6} g_{\rm PS} f_a   
\, .
\eeq
Hence, using the standard relation between axion decay constant and mass \cite{Gorghetto:2018ocs}
\beq 
\label{eq:axionmass}
m_a = 5.691(51) \( \frac{10^{12} \ \text{GeV}}{ f_a} \) \, \text{$\mu$eV} \, , 
\eeq 
we can link the axion mass to $M_{\rm PS}$, via 
\beq
\label{eq:mavsMPS}
\boxed{
m_a \approx 2.788\( \frac{10^{13} \ \text{GeV}}{M_{\rm PS}} \) g_{\rm PS} \ \text{$\mu$eV} 
\, .
}
\eeq
In the next Section we will constrain $(M_{\rm PS}, g_{\rm PS})$, and in turn the axion mass, 
via a RG analysis of (partial) gauge coupling unification in Pati-Salam.

\subsection{Renormalization group analysis} 
\label{sec:RGE}

Let us consider the breaking pattern 
\beq
\G_{\rm PS} \times \U(1)_{\rm PQ} \xrightarrow[M_{\rm PS}]{\vev{\S}} \G^{(\slashed{\P})}_{\rm LR} \xrightarrow[M_{\rm LR}]{\vev{\Delta_R}}  \G_{\rm SM} \xrightarrow[M_{Z}]{\langle \Phi_{1,15}\rangle} \SU(3)_C \times \U(1)_{\rm EM}
\, , 
\eeq
where $M_{\rm PS}$, $M_{\rm LR}$ and $M_{Z}$ denote the renormalization scales associated with 
$\G_{\rm PS}$, $\G^{(\slashed{\P})}_{\rm LR}$ and $\G_{\rm SM}$, respectively. 
In order to determine the beta-functions which govern the RG evolution in the 
two running steps, we assume that 
the scalar spectrum obeys the so-called extended survival hypothesis (ESH) \cite{delAguila:1980qag} 
which requires that at every stage of the symmetry breaking chain only those scalars are present that develop 
a VEV at the current or the subsequent levels of the spontaneous symmetry breaking. 
The ESH is equivalent to the requirement of the minimal number of fine-tunings to be imposed onto the scalar 
potential \cite{Mohapatra:1982aq}. The surviving scalars at the $M_{Z}$ and $M_{\rm LR}$ scales are 
displayed in \Table{tab:ESH} while the corresponding one- and two-loop beta coefficients are collected in 
\eqs{eq:bfSM}{eq:bfLRNoP}. When the $\P$ symmetry is broken at the $M_{\rm LR}$ scale 
($\G_{\rm LR}$ case) an extra left-handed triplet $\subset \Delta_L$ automatically accompanies (without extra fine-tunings) 
the Left-Right symmetry breaking right-handed triplet $\subset \Delta_R$, resulting in a different 
$M_{\rm LR} \to M_{\rm PS}$ RG evolution compared to the $\G^{\slashed{P}}_{\rm LR}$ case. 
\begin{table}[htp]
\begin{center}
\begin{tabular}{|c|c|c|}
\hline
 & $\G_{\rm SM}: \ M_Z \to M_{\rm LR}$ running & $\G^{(\slashed{\P})}_{\rm LR}: \ M_{\rm LR} \to M_{\rm PS}$ running \\
\hline
$\P$ & $(1,2,\tfrac{1}{2}) \subset \{\Phi_1,\Phi_{15} \}$ & 
\makecell{$(1,2,2,0) \subset \{\Phi_1,\Phi_{15} \}$ \\ $(1,1,3,1) \subset \Delta_R$ \\ $(1,3,1,1) \subset \Delta_L$} \\ 
\hline
$\slashed{\P}$ & $(1,2,\tfrac{1}{2}) \subset \{\Phi_1,\Phi_{15} \}$ & 
\makecell{$(1,2,2,0) \subset \{\Phi_1,\Phi_{15} \}$ \\ $(1,1,3,1) \subset \Delta_R$} \\
\hline
\end{tabular}
\end{center}
\caption{Surviving intermediated-scale scalars and their Pati-Salam 
origin, according to the ESH. 
Transformation properties 
under the $\G_{\rm SM}$ and $\G^{(\slashed{\P})}_{\rm LR}$ groups
are given explicitly.
}
\label{tab:ESH}
\end{table}

We first focus on a one-loop RG analysis, in order to grasp an analytical understanding of the 
correlation among mass scales. Starting with the electroweak values of the three SM gauge couplings 
\cite{Mihaila:2012pz}
\begin{align}
\alpha_1 (M_Z) &= 0.016923 \pm 0.000004 \, , \\
\alpha_L (M_Z) &= 0.03374 \pm 0.00002 \, , \\ 
\alpha_C (M_Z) &= 0.1173 \pm 0.0007 \, ,
\end{align}
(these data refer to the modified minimal subtraction scheme ($\overline{\rm MS}$) in the full SM, 
i.e.~the top being not integrated out)
at the $M_Z = 91.1876$ GeV scale, 
where $\alpha_i \equiv g^2_i / (4 \pi)$ and $g_1 = \sqrt{\frac{5}{3}} g_Y$ is the GUT-normalized hypercharge coupling.    
The SM gauge couplings are evolved up to $M_{\rm LR}$ with one-loop SM beta functions 
\begin{align}
\label{SMrun1}
\alpha_1^{-1} (M_{\rm LR}) &= \alpha_1^{-1} (M_Z) - \frac{a_1^{\rm SM}}{2 \pi} \log\frac{M_{\rm LR}}{M_Z} \, , \\
\label{SMrunL}
\alpha_L^{-1} (M_{\rm LR}) &= \alpha_L^{-1} (M_Z) - \frac{a_L^{\rm SM}}{2 \pi} \log\frac{M_{\rm LR}}{M_Z} \, , \\
\label{SMrunC}
\alpha_C^{-1} (M_{\rm LR}) &= \alpha_C^{-1} (M_Z) - \frac{a_C^{\rm SM}}{2 \pi} \log\frac{M_{\rm LR}}{M_Z} \, , 
\end{align}
with the SM beta coefficients given in \eq{eq:bfSM}. 
The tree-level matching of the gauge couplings at $M_{\rm LR}$ is given by 
\begin{align}
\label{mathcPS1}
\alpha_1^{-1} (M_{\rm LR}) &= \frac{3}{5} \alpha_R^{-1} (M_{\rm LR}) + \frac{2}{5} \alpha_{B-L}^{-1} (M_{\rm LR}) \, , \\
\label{mathcPSL}
\alpha_L (M_{\rm LR}) &= \alpha_L (M_{\rm LR}) \, , \\
\label{mathcPSC}
\alpha_C (M_{\rm LR}) &= \alpha_{C} (M_{\rm LR}) \, ,  
\end{align}
where \eq{mathcPS1} comes from the relation
\beq 
Q_Y = \sqrt{\frac{3}{5}} Q_R + \sqrt{\frac{2}{5}} Q_{B-L} \, , 
\eeq
between the properly normalized generators $Q_Y = \sqrt{\frac{3}{5}} Y$, $Q_R = T^3_{R}$ and $Q_{B-L} = \sqrt{\frac{3}{2}} \frac{B-L}{2}$.
The second stage of running between $M_{\rm LR}$ and $M_{\rm PS}$ is given by 
\begin{align}
\label{LRrunBmL}
\alpha_{B-L}^{-1} (M_{\rm PS}) &= \alpha_{B-L}^{-1} (M_{\rm LR}) - \frac{a_{B-L}^{\rm LR}}{2 \pi} \log\frac{M_{\rm PS}}{M_{\rm LR}} \, , \\
\label{LRrunL}
\alpha_L^{-1} (M_{\rm PS}) &= \alpha_L^{-1} (M_{\rm LR}) - \frac{a_L^{\rm LR}}{2 \pi} \log\frac{M_{\rm PS}}{M_{\rm LR}} \, , \\
\label{LRrunR}
\alpha_R^{-1} (M_{\rm PS}) &= \alpha_R^{-1} (M_{\rm LR}) - \frac{a_R^{\rm LR}}{2 \pi} \log\frac{M_{\rm PS}}{M_{\rm LR}} \, , \\
\label{LRrunC}
\alpha_C^{-1} (M_{\rm PS}) &= \alpha_C^{-1} (M_{\rm LR}) - \frac{a_C^{\rm LR}}{2 \pi} \log\frac{M_{\rm PS}}{M_{\rm LR}} \, . 
\end{align}
The value of the one-loop beta coefficients are given in \eqs{eq:bfLRP}{eq:bfLRNoP}. 
Finally, the tree-level matching of the gauge couplings at the $M_{\rm PS}$ scale is 
\begin{align}
\label{mathcPS1a}
\alpha_{B-L} (M_{\rm PS}) &= \alpha_{C} (M_{\rm PS}) = \alpha_{\rm PS} (M_{\rm PS}) \, , \\
\label{mathcPSLb}
\alpha_R (M_{\rm PS}) &= \alpha_L (M_{\rm PS}) = \alpha_{\rm LR} (M_{\rm PS}) \, ,
\end{align}
which correspond to the case where $\P$ is restored only at the $M_{\rm PS}$ scale 
($\G^{\slashed{\P}}_{\rm LR}$), 
while the case when $\P$ is restored already at $M_{\rm LR}$ 
($\G_{\rm LR}$) is obtained by setting 
$\alpha_L (M_{\rm LR}) = \alpha_R (M_{\rm LR})$ and $a_L^{\rm LR} = a_R^{\rm LR}$. 

\begin{figure}[t]
\centering
\includegraphics[width=10cm]{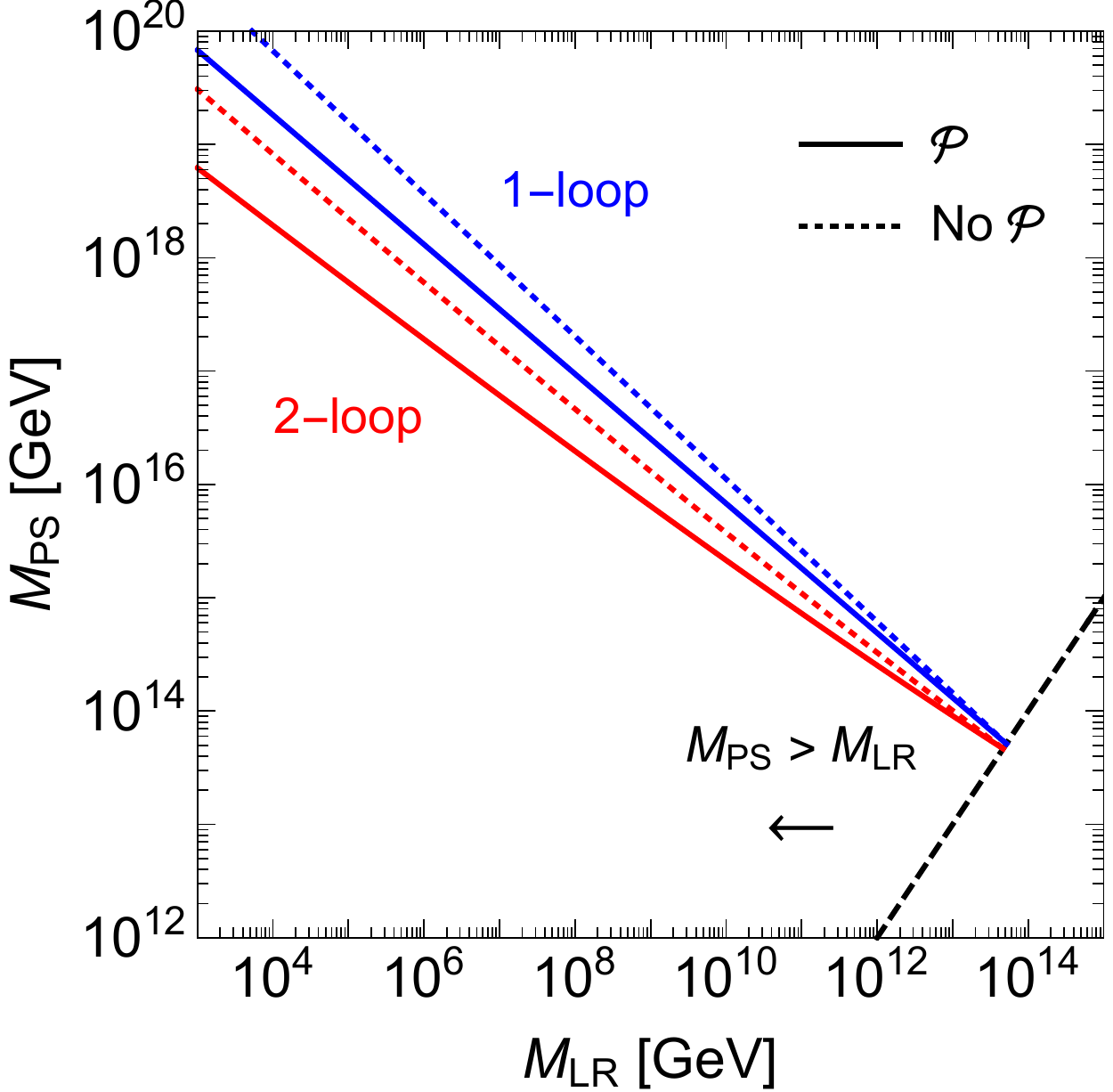} \quad
\caption{Pati-Salam unification scale 
as a function of the Left-Right symmetry breaking scale,  
from a one-loop (blue lines) and a two-loops (red lines) RG analysis.    
The full (dotted) lines correspond to the case where $\P$ is broken at the $M_{\rm LR}$ ($M_{\rm PS}$) scale.   
}
\label{fig:MPSvsMLR}       
\end{figure}

\begin{figure}[th]
\centering
\includegraphics[width=7cm]{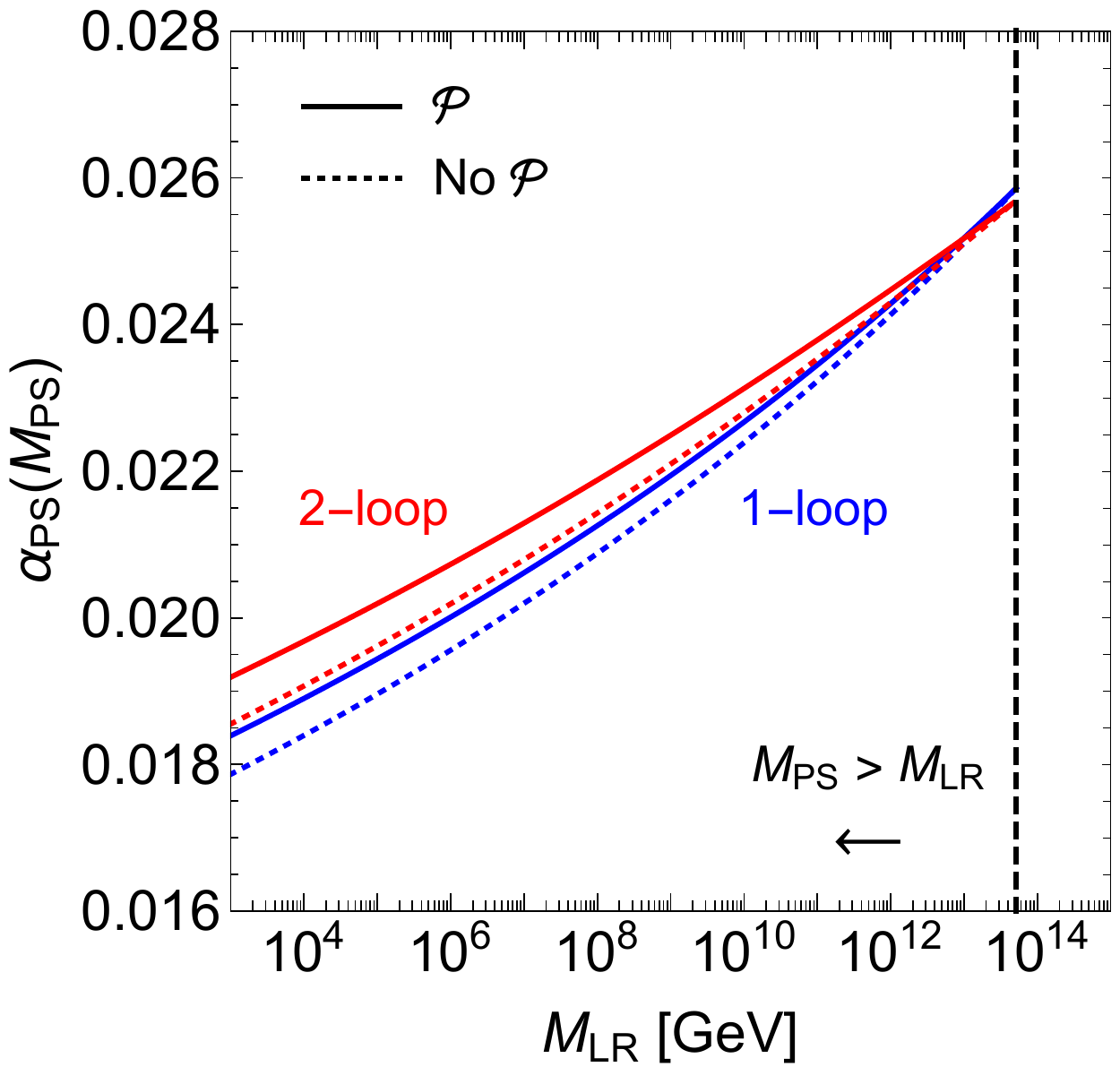} \quad
\includegraphics[width=7cm]{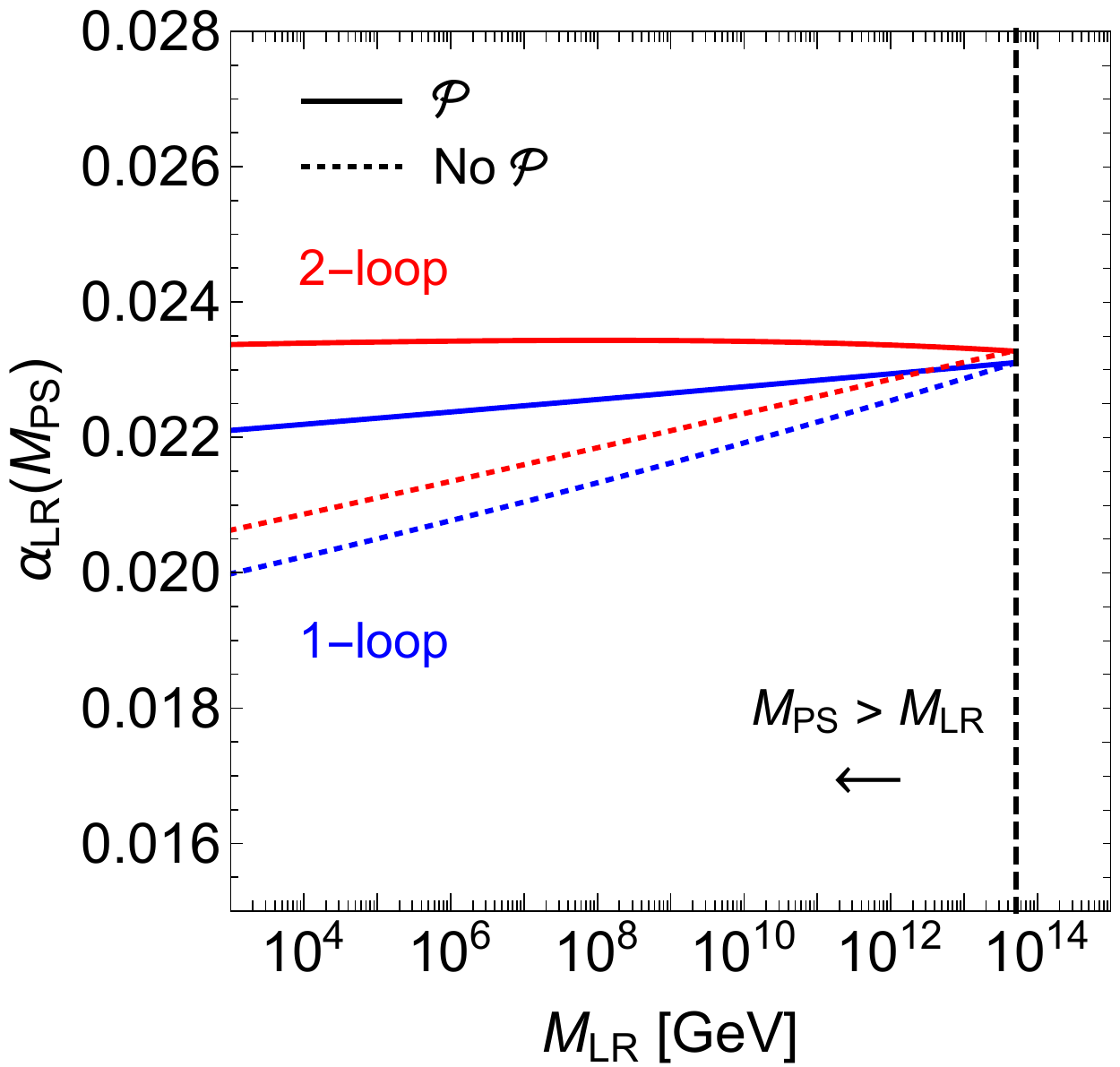}
\caption{
Pati-Salam gauge coupling (left panel) and Left-Right gauge coupling (right panel) 
at the Pati-Salam unification scale, as a function of the Left-Right symmetry breaking scale. 
}
\label{fig:alphasvsMLR}       
\end{figure}

Reshuffling the RG equations and the matching conditions above one obtains
\beq
\label{eq:MLRMPS1l}
\( \frac{M_{\rm PS}}{M_{\rm LR}} \)^{\frac{\Delta a^{\rm LR}}{2\pi}} 
\( \frac{M_{\rm LR}}{M_{\rm Z}} \)^{\frac{\Delta a^{\rm SM}}{2\pi}} 
= \exp{\(\alpha_1^{-1} (M_Z) - \frac{3}{5} \alpha_L^{-1} (M_Z) - \frac{2}{5} \alpha_C^{-1} (M_Z) \)} \, , 
\eeq
with
\begin{align}
\label{eq:defaSM}
\Delta a^{\rm SM} &= a_1^{\rm SM} - \frac{3}{5} a_L^{\rm SM} - \frac{2}{5} a_C^{\rm SM} \, , \\
\label{eq:defaLR}
\Delta a^{\rm LR} &= 
\frac{3}{5} \( a_R^{\rm LR} - a_L^{\rm LR} \) 
+\frac{2}{5} \( a_{B-L}^{\rm LR} - a_C^{\rm LR} \) \, , 
\end{align}
which allows to solve for $M_{\rm PS}$ as a function of $M_{\rm LR}$ 
and the known SM parameters at the electroweak scale.  
In particular, using the one-loop beta coefficients in 
\app{app:2loopbf} we have 
$\Delta a^{\rm SM} = \frac{44}{5} $ and $\Delta a^{\rm LR} =\frac{28}{5}$ ($\Delta a^{\rm LR} = \frac{27}{5}$) for the 
$\G_{\rm LR}$ ($\G^\slashed{P}_{\rm LR}$) case. 
Hence, from \eq{eq:defaSM} we readily conclude that $M_{\rm PS}$ is a decreasing function of $M_{\rm LR}$. 
Similarly, we can solve for the $\G_{\rm PS}$ gauge couplings at the $M_{\rm PS}$ and obtain
\begin{align}
\label{eq:alphaPS1l}
\alpha_{\rm PS}^{-1} (M_{\rm PS}) &= 
\alpha_C^{-1} (M_Z) 
- \frac{a_C^{\rm SM}}{2 \pi} \log\frac{M_{\rm LR}}{M_Z} 
- \frac{a_C^{\rm LR}}{2 \pi} \log\frac{M_{\rm PS}}{M_{\rm LR}}
\, , \\
\label{eq:alphaLR1l}
\alpha_{\rm LR}^{-1} (M_{\rm PS}) &= 
\alpha_L^{-1} (M_Z) 
- \frac{a_L^{\rm SM}}{2 \pi} \log\frac{M_{\rm LR}}{M_Z} 
- \frac{a_L^{\rm LR}}{2 \pi} \log\frac{M_{\rm PS}}{M_{\rm LR}}
\, . 
\end{align}
The dependence of Pati-Salam breaking scale from $M_{\rm LR}$, 
as well as that of the gauge couplings at  
$M_{\rm PS}$ 
is displayed respectively in \fig{fig:MPSvsMLR} and \ref{fig:alphasvsMLR} 
for the physical region $M_{\rm PS} > M_{\rm LR}$, 
where we also included the results of a two-loop RG analysis, whose details are described in \app{app:2loopbf}.
As already anticipated, the Pati-Salam unification scale is a decreasing function of the Left-Right 
symmetry breaking scale. Two-loop effects are especially relevant for a low-scale $M_{\rm LR}$ 
and they tend to lower $M_{\rm PS}$ by up to one order of magnitude 
(or, equivalently, for fixed $M_{\rm PS}$ to lower $M_{\rm LR}$ 
by up to two orders of magnitude). 
The case of $\G_{\rm LR}$ (with $\P$ broken at $M_{\rm LR}$) systematically leads to lower values of 
$M_{\rm PS}$ and $M_{\rm LR}$ 
compared to the case of $\G^{\slashed{P}}_{\rm LR}$ (with $\P$ broken at $M_{\rm PS}$). 

A word of caution is in order here about missing scalar threshold 
corrections. Particles from the scalar spectrum might sizeably change the results 
of this analysis, if they are not clustered around the mass scale 
of the massive vector bosons, which are identified with the 
renormalization scales at which the matching is performed 
(for a more precise definition see \eq{eq:1loopmatchgen}). 
To improve on this point one should take into account the constraints 
coming form the minimization of the full 
$\G_{\rm PS} \times \U(1)_{\rm PQ}$
scalar potential, in order to 
obtain the range of variation of the scalar thresholds 
allowed by the vacuum manifold. It goes without saying that this is 
a highly non-trivial task. As a partial justification of the ESH $=$ minimal fine-tuning 
hypothesis \cite{delAguila:1980qag,Mohapatra:1982aq}, 
we note that strong violations of the latter would be difficult to be reconciled 
with the idea that gauge hierarchies could
arise due to environmental selection/cosmological evolution 
(see also footnote (\ref{foot:gaugeH})).


\begin{figure}[th]
\centering
\includegraphics[width=14cm]{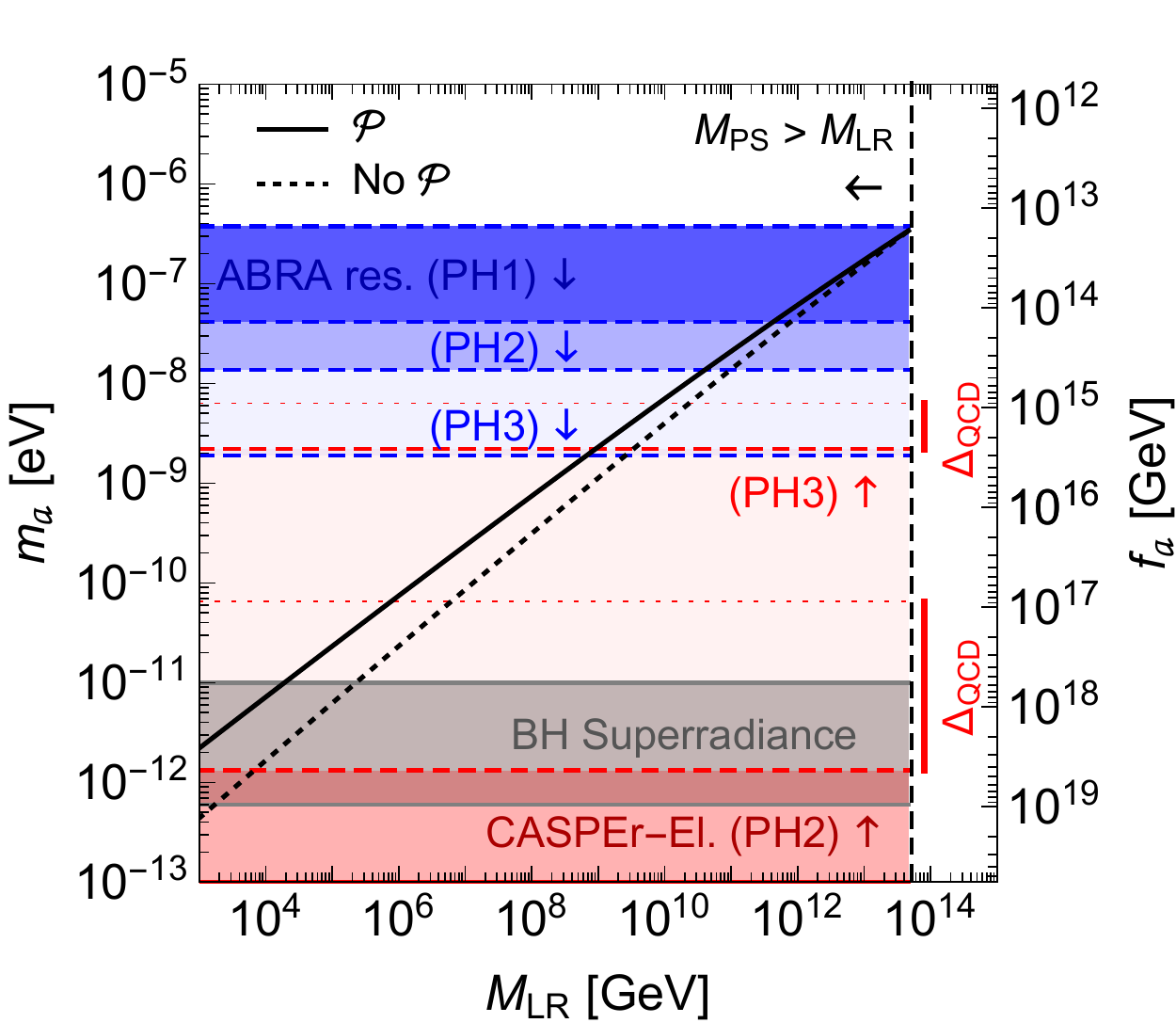} 
\caption{Axion mass dependence from the Left-Right symmetry breaking scale
(from two-loop RG analysis). 
The full (dotted) black line corresponds to the case where $\P$ is broken at the $M_{\rm LR}$ ($M_{\rm PS}$) scale.  
The current bound from Black Hole Superradiance (gray) and the future reach 
in different phases of 
ABRACADABRA (blue) and CASPEr-Electric (red) are shown as well (see text for details).} 
\label{fig:maxionvsMLR}       
\end{figure}

We can now proceed with the main goal of the RG analysis, namely to express the axion mass 
as a function of $M_{\rm LR}$, via the relation in \eq{eq:mavsMPS}.  
This correlation is shown in \fig{fig:maxionvsMLR}, where we report directly the two-loop result 
for the two cases $\G_{\rm LR}$ ($\P$) and $\G^{\slashed{\P}}_{\rm LR}$ (No $\P$), 
together with the current bounds from Black Hole Superradiance and the experimental prospects 
of future axion DM experiments, whose phenomenological 
implications are described in the next Section.

\clearpage

\section{Phenomenology} 
\label{sec:pheno}

From the two-loop RG analysis
we have inferred the mass windows displayed in \Table{tab:twoloopscales}, 
where the lower bounds take already into account the exclusion limit in \eq{eq:exclBHS}. 
In particular, the correlations among the mass scales can be read off 
\fig{fig:MPSvsMLR} ($M_{\rm PS}$ vs.~$M_{\rm LR}$) and \fig{fig:maxionvsMLR} 
($m_a$ vs.~$M_{\rm LR}$).

\begin{table}[htp]
\begin{center}
\begin{tabular}{|c|c|c|c|}
\hline
& $m_a$ [eV] & $M_{\rm LR}$ [GeV] & $M_{\rm PS}$ [GeV] \\
\hline
$\G_{\rm LR}$ & $[10^{-11}, 3.4 \times 10^{-7}]$ & $[2.0 \times 10^{4}, 4.7 \times 10^{13}]$ 
& $[1.4 \times 10^{18}, 4.7 \times 10^{13}]$ \\
\hline 
$\G^{\slashed{\P}}_{\rm LR}$ & $[10^{-11}, 3.4 \times 10^{-7}]$ & $[2.3 \times 10^{5}, 4.7 \times 10^{13}]$
& $[1.4 \times 10^{18}, 4.7 \times 10^{13}]$ \\
\hline
\end{tabular}
\end{center}
\caption{Mass windows from two-loop RG analysis.}
\label{tab:twoloopscales}
\end{table}

In this Section we discuss the phenomenological implications of 
those mass ranges,  
both from the point of view of axion physics
and the Pati-Salam/Left-Right symmetry breaking scales.  

\subsection{Cosmological and astrophysical constraints}
\label{sec:astrocosmo}

Let us start by addressing some relevant cosmological and astrophysical constraints.  
Very light axion DM tends to be overproduced via the misalignment 
mechanism \cite{Preskill:1982cy,Abbott:1982af,Dine:1982ah}, and 
the measured amount of cold DM  can only be explained if the PQ symmetry remained broken 
during inflation and never restored after it, 
which corresponds to the so-called pre-inflationary PQ breaking scenario 
(more precisely, the post-inflationary PQ breaking scenario is excluded for $m_a \lesssim 30$ $\mu$eV \cite{Borsanyi:2016ksw}). 
A late inflationary phase $H_I \lesssim M_{\rm PS} \approx f_a$ (cf.~\eq{eq:MPSfa})
is 
also supported by other cosmological issues of the Pati-Salam axion 
framework related to the formation of topologically stable defects, 
which tend to dominate the energy density of the Universe, unless inflated away. 
These include: 
$i)$ Magnetic monopoles from Pati-Salam breaking;
$ii)$ Domain walls from spontaneous breaking of $\P$; 
$iii)$ Axion domain walls 
at the QCD phase transition. 
While the domain wall problems are not specific of Pati-Salam and 
have been widely discussed in the literature (for a review see e.g.~\cite{Vilenkin:1984ib}), 
we dwell a bit on the less known physics of Pati-Salam monopoles in \sect{sec:PSmonopoles}. 
After that we discuss other constraints related to axion DM (\sect{sec:PSmonopoles}) 
and Black Hole Superradiance (\sect{sec:BHsuper}). 

\subsubsection{Pati-Salam monopoles}
\label{sec:PSmonopoles}

Pati-Salam monopoles are topologically stable scalar-gauge field configuration 
arising from $\G_{\rm PS} \to \G_{\rm SM}$ breaking, with magnetic charge 
$Q_{\rm mag} = 4\pi / e$ and mass $\M_{\rm PS} \sim M_{\rm PS} / \alpha_{\rm PS}$.  
Although they were originally investigated in the context 
of intermediate symmetry breaking stages of SO(10) \cite{Lazarides:1980va,Lazarides:1980cc,Dawson:1982sc}, 
they have some distinct features with respect to GUT monopoles 
which make them interesting by their own. 
 
A flux of Pati-Salam monopoles hitting an Earth-based detector could 
actually lead to a spectacular signature, 
since according to Sen \cite{Sen:1984ku,Sen:1984qw}
they are expected to catalyze $\Delta (B+L) = 3$ 
violating processes via the weak `t Hooft anomaly 
with a geometrical 
cross 
section, i.e.~not suppressed by the Pati-Salam breaking scale 
or by other non-perturbative factors. 
The conservative Kibble estimate \cite{Kibble:1976sj}  
of one monopole per Hubble horizon 
provides a lower bound on today's monopole number density (normalized to entropy density)\footnote{In fact, 
a refined estimate from Zurek \cite{Zurek:1985qw}, which takes into account the 
timescale 
of the phase transition, leads to a substantially larger abundance than the original Kibble estimate, 
especially in the presence of a second-order phase transition \cite{Murayama:2009nj,Khoze:2014woa}.}
\beq 
\label{eq:novsKibble}
\frac{n_{\rm PS}}{s} \gtrsim \( \frac{T_{\rm PS}}{M_{\rm Pl}} \)^3 \, ,
\eeq
where $T_{\rm PS} \approx M_{\rm PS} \gtrsim 10^{13}$ GeV 
is the critical temperature of the Pati-Salam phase transition.  
Hence, for the lowest value allowed 
by the RG analysis, $M_{\rm PS} = 4.7 \times 10^{13}$ GeV 
(cf.~\Table{tab:twoloopscales}) one has $n_{\rm PS} / s \gtrsim 5.7 \times 10^{-17}$, 
which overshoots the critical energy density of the Universe\footnote{The 
relic abundance is related to the 
number density via 
$\Omega_{\rm PS} h^2 = 2.8 \times 10^{8} \( n_{\rm PS} / s \) \( \M_{\rm PS} / \text{GeV} \)$.} 
and is also well above the indirect Parker's bound \cite{Parker:1970xv,Turner:1982ag}, $n_{\rm PS} / s \lesssim 10^{-26}$,  
and the direct detection limits from the MACRO Collaboration \cite{Ambrosio:2002qq}, 
$n_{\rm PS} / s \lesssim 2 \times 10^{-28}$. On the other hand, 
differently from GUT monopoles, Pati-Salam monopoles 
are not subject to the much more stringent bounds \cite{Ueno:2012md} 
from the catalysis of proton decay 
due to the Callan-Rubakov effect \cite{Callan:1982au,Rubakov:1982fp}.

Although in the model considered in this paper Pati-Salam monopoles need to be inflated away, 
we note that the observational window 
of two orders of magnitude between the Parker's bound and direct detection limits 
might be populated in models 
with $M_{\rm PS} \ll 10^{13}$ GeV, e.g.~in the ballpark of $M_{\rm PS} \approx 10^{10}$ GeV according to the 
naive Kibble estimate.

\subsubsection{Axion relic density and iso-curvature bounds}
\label{sec:axionDMbounds}

In the pre-inflationary PQ breaking scenario, 
the axion's relic abundance depends both on the mass and on the initial value of the axion field $a_i$ in units of the decay constant, $\theta_i = a_i/f_a$,  
inside the causally connected region which is inflated into our visible Universe, cf.~\cite{Borsanyi:2016ksw,Ballesteros:2016xej}:
\begin{equation}
\label{eq:axionDMrelic}
\Omega_ah^2 
= 0.12\,\left({ 3.4 \times 10^{-7}~{\rm eV}\over m_a}\right)^{1.165}\,
\left(\frac{\theta_i}{0.13}\right)^2  \,, 
\end{equation}
where we have normalized the axion mass to the upper bound in \Table{tab:twoloopscales} 
coming from the RG analysis. Thus an axion close to that boundary can reproduce the whole 
DM, without the need of tuning the initial misalignment angle. 
In this cosmological scenario, however, 
iso-curvature quantum fluctuations of a massless axion field during inflation may leave an imprint in the temperature fluctuations of the cosmic microwave background \cite{Linde:1985yf,Seckel:1985tj}, whose amplitude is stringently constrained by observations. In the case that the $\S$ field hosting the axion 
stays at the broken minimum of the potential throughout inflation 
(i.e.~for the inflaton field not residing in $\S$), those constraints translate into an upper bound 
on the Hubble expansion rate during inflation \cite{Beltran:2006sq,Hertzberg:2008wr,Hamann:2009yf} 
\begin{equation}
H_I < 9.8 \times 10^7\,{\rm GeV} \left( \frac{3.4 \times 10^{-7}\,\text{eV}}{m_a}\right)^{0.4175}
\,,
\end{equation}
which is consistent with a late inflationary phase in order to 
dilute the cosmic density of monopoles and domain walls. 
%

\subsubsection{Black Hole Superradiance}
\label{sec:BHsuper}

Although super-light axions are free from standard astrophysical bounds 
due to stellar evolution, they are subject to limits from Black Hole Superradiance 
as long as the axion decay constant approaches the Planck mass.  
In fact, axions can form gravitational bound states around black holes whenever their 
Compton wavelength is of the order of the black holes radii. 
The phenomenon of superradiance \cite{Penrose:1969pc}
then guarantees that the axion occupation numbers grow exponentially, 
providing a way to extract very efficiently energy and angular momentum from the black hole 
\cite{Arvanitaki:2010sy,Arvanitaki:2014wva}. 
The rate at which the angular momentum is extracted depends on the black hole mass and so the presence of axions could be inferred by observations of black hole masses and spins. 
A recent analysis excludes the mass window \cite{Cardoso:2018tly} 
\beq 
\label{eq:exclBHS}
m_a \in [6 \times 10^{-13} , 10^{-11} ] \ \text{eV} \, , 
\eeq
which is shown as a gray band in \fig{fig:maxionvsMLR}. 
It should be stressed that the Black Hole Superradiance bound 
does not assume the axion being DM and it just relies on the 
universal axion coupling to gravity through its mass.

\subsection{Axion Dark Matter experiments}
\label{eq:axionDM}

Axion DM experiments turn out to be the most 
powerful probes of the Pati-Salam axion 
(under the assumption that the axion comprises the whole DM), 
since they can cover the whole mass window 
$m_a \in [10^{-11}, 3.4 \times 10^{-7}]$ eV 
predicted by the RG analysis (cf.~\Table{tab:twoloopscales}). 
In particular, this is possible due to the complementarity 
of ABRACADABRA \cite{Kahn:2016aff} and 
CASPEr-Electric \cite{Budker:2013hfa,JacksonKimball:2017elr}, 
which probe the axion mass parameter space region from opposite directions 
(cf.~\fig{fig:maxionvsMLR}).

\subsubsection{ABRACADABRA}

The axion DM experiment ABRACADABRA \cite{Kahn:2016aff} 
has very good prospects to probe the 
axion-photon coupling 
for masses $m_a \lesssim 4 \times 10^{-7}$ eV. Such low values are 
notoriously difficult to be reached for standard cavity experiments 
due to the need of matching axion wavelengths with extremely large cavity sizes $\gtrsim 50$ m. 
ABRACADABRA uses instead a different detection concept 
based on a toroidal magnet and a pickup loop to detect the variable magnetic flux induced by the oscillating current produced by DM axions in the static (lab) magnetic field. 
The experiment can operate either in broadband or resonant modus 
by using an untuned or a tuned magnetometer respectively.  
A small scale prototype ABRACADABRA-10 cm \cite{Ouellet:2018beu} 
has already given exclusion limits competitive with astrophysics 
for axion masses $\in [3.1 \times 10^{-10}, 8.3 \times 10^{-9}]$ eV. 
The projected sensitivities (from \cite{Kahn:2016aff}) on the axion mass
are displayed in \fig{fig:maxionvsMLR} via blue bands.  
They refer to three different phases of the experiment 
in the resonant approach, 
which is more sensitive to the standard QCD axion region,  
and they also assume $E/N = 8/3$,  
which applies to the Pati-Salam axion considered in this work (cf.~\eq{eq:PSaxioncoupl}) 
and in general to any GUT axion model. 

\subsubsection{CASPEr}

CASPEr-Electric \cite{Budker:2013hfa,JacksonKimball:2017elr} 
employs nuclear magnetic resonance techniques to search 
for an oscillating nEDM \cite{Graham:2013gfa}
\beq 
d_n (t) = g_{d}\,\frac{\sqrt{2\rho_{\text{DM}}}}{m_a} \cos (m_a\,t) \, ,
\eeq
where $g_{d} = C_{an\gamma} / (m_n f_a)$ 
is the model-independent coupling of the axion to the nEDM operator 
defined in \eq{eq:Laint1} 
and $\rho_{\rm DM} \approx 0.4\ \text{GeV}/\text{cm}^3$ is the local energy density of axion DM. 
The numerical value of $C_{an\gamma}$ (cf.~\eq{eq:Cangamma}) takes over the static nEDM 
calculation of \cite{Pospelov:1999mv} based on QCD sum rules and amounts to a theoretical error of about 
$40 \%$. 
In \fig{fig:maxionvsMLR} we show in red bands 
the axion mass reach of CASPEr-Electric \cite{JacksonKimball:2017elr} 
for phases 2 and 3, including as well on the right side of the plot 
the size of the QCD error (denoted by $\Delta_{\rm QCD}$). 
Remarkably, the QCD error turns out to be strongly correlated with the projected sensitivities 
shown in Ref.~\cite{JacksonKimball:2017elr}, so that a future reduction of the theoretical 
error on the static nEDM 
(e.g.~via Lattice QCD techniques \cite{Abramczyk:2017oxr,Dragos:2019oxn}) 
might have a non-trivial impact on the sensitivity reach of CASPEr-Electric. 

On the other hand, 
the projected sensitivity of CASPEr-Wind \cite{JacksonKimball:2017elr}, which  exploits the axion nucleon 
($N=p,n$) coupling $g_{aN}=C_{aN}/(2f_a)$ (with $C_{aN}$ given in \eqs{eq:Cap}{eq:Can}) 
to search for an axion DM wind due to the movement of the Earth through the 
Galactic DM halo \cite{Graham:2013gfa}, 
misses the preferred coupling vs.~mass region by 
at least two orders of magnitude, even in its phase 2.

\subsection{Pati-Salam signatures}
\label{eq:PSLRsignatures}

Given the lower bound on the Pati-Salam breaking scale inferred from the RG analysis, 
$M_{\rm PS} \gtrsim 10^{13}$, 
genuine signatures of Pati-Salam dynamics turn out to be 
difficult to be experimentally accessible, 
as briefly reviewed in the following. 

\subsubsection{Rare meson decays}

The vector leptoquark $\X_\mu$ mediates tree-level rare meson decays 
$K_L,\, B^0, B_s^0 \to \ell_i \, \ell_j$ (for $\ell = (e,\mu,\tau)$), 
which turn out to be loop- and chirally-enhanced with respect to the SM 
contribution, and hence provide a powerful flavour probe of the 
Pati-Salam model \cite{Valencia:1994cj,Kuznetsov:1994tt}. 
A recent collection of bounds which takes also into account flavour mixing can be found in 
Ref.~\cite{Smirnov:2018ske}. 
For instance, for maximal mixing the most sensitive channel is 
$K_L \to \mu e$, which probes leptoquark masses up to $10^6$ GeV, 
hence still much below the Pati-Salam breaking scale emerging from the RG analysis. 

\subsubsection{Baryon number violation}


A standard signature of Pati-Salam dynamics are 
$\Delta B = 2$ processes, in particular $n$-$\bar{n}$ oscillations \cite{Mohapatra:1980qe},  
which are described by $d=9$ SM operators of the type
\beq 
\label{eq:opnantin}
\frac{1}{\Lambda^5_{\Delta B=2}} \, 
\[ u_R d_R d_R u_R d_R d_R + 
q_L q_L q_L q_L d_R d_R \]
\, .  
\eeq
Present bounds on nuclear instability and direct reactor oscillations experiments 
yield bounds at the level of $\Lambda_{\Delta B=2} \gtrsim 100$ TeV \cite{Mohapatra:2009wp}. 
In the present Pati-Salam$\times \U(1)_{\rm PQ}$ model (see also \cite{Saad:2017pqj})
only operators of the type $q_L q_L q_L q_L d_R d_R$ are generated,  
which are mediated by color sextet scalar di-quark fields $\Delta_{qq}$ contained in 
the Pati-Salam representations $\Delta_{R} \sim (10,1,3)$ and $\Delta_{L} \sim (10,3,1)$, with 
strength (schematically)
\beq 
\label{eq:strengthDeltaB2}
\frac{1}{\Lambda^5_{\Delta B=2}} \sim \frac{\eta \, y \, v_{R}}{m^6_{\Delta_{qq}}} \, ,
\eeq 
where $\eta$ and $y$ denote respectively the 
scalar coupling of $\Delta_R^2 \Delta_L^2$ (cf.~\eq{eq:VPQ}) 
and the Yukawa coupling of $\Delta_{qq}$ to quarks (cf.~\eq{eq:LYPS}). 
Hence, $n$-$\bar{n}$ oscillations could be visible only if those color sextets were unnaturally light 
$m_{\Delta_{qq}} \ll M_{\rm PS}$. 

Nucleon decay from the scalar sector of Pati-Salam is possible as well, 
as originally observed in \cite{Pati:1983zp,Pati:1983jk}.  
In the presence of the Higgs multiplet $\Phi_{15} \sim (15,2,2)$, required by a realistic fit 
to SM fermion masses, 
the spontaneous breaking of $\U(1)_{B-L}$ can lead to 
$B+L$ preserving nucleon decay modes of the type 
$N \to \ell + \text{meson}$ or even three-lepton 
decay modes $N \to \ell + \bar{\ell} + \ell$ \cite{ODonnell:1993kdg},  
associated respectively to $d=9$ and $d=10$ SM effective operators, 
whose short-distance origin can be traced back into the scalar potential couplings 
$\gamma_{15}$ and $\delta_{15}$ 
in \eq{eq:VPQ} (see \cite{Saad:2017pqj} for a more detailed account). 
As in the case of $n$-$\bar{n}$ oscillations, for these exotic nucleon decay modes 
to be observable, one or more scalar fragments of $\Phi_{15}$ and $\Delta_{R,L}$ 
mediating those nucleon decay operators need to be $\lesssim 100$ TeV, 
unnaturally light compared to the Pati-Salam breaking scale. 

\subsection{Left-Right signatures}
\label{eq:LRonlysignatures}

A sliding Left-Right symmetry breaking scale 
$M_{\rm LR} \in [2.0 \times 10^{4}, 4.7 \times 10^{13}]$ GeV 
(e.g.~in the case of $\P$ broken at the $M_{\rm LR}$ scale, cf.~\Table{tab:twoloopscales}) 
offers potentially observables signatures related to the Left-Right breaking dynamics, 
together with the possibility of observing a correlated signal with axion physics.
In the following, we 
consider two opposite scenarios in which the Left-Right symmetry is broken either at high or low scales.

\subsubsection{High-scale Left-Right breaking}

The high-scale Left-Right breaking scenario, corresponding to a 
single step breaking with $M_{\rm LR} \approx M_{\rm PS} \approx 4.7 \times 10^{13}$ GeV is 
motivated by naturalness arguments. 
In fact, it simultaneously minimizes the parameter space tuning 
in three sectors of the theory: 
$i)$ it avoids the extra\footnote{\label{foot:gaugeH}The $(M_Z / M_{\rm PS})^2$ tuning of the 
electroweak scale cannot be avoided. 
It is conceivable that the solution of the latter problem does not rely on a stabilizing symmetry.  
For instance, a light Higgs might be arise as an attractor point in  
the cosmological evolution of the Universe \cite{Dvali:2003br,Dvali:2004tma}.} 
$(M_{\rm LR} / M_{\rm PS})^2$ tuning of triplet fields (cf.~\Table{tab:ESH});   
$ii)$ it reduces the tuning of the initial axion misalignment angle in order not to overshoot the 
axion DM relic density (cf.~\eq{eq:axionDMrelic}); 
$iii)$ it mitigates the tuning in the Dirac neutrino mass matrix, which turns out to be strongly correlated with the 
up-quark mass matrix (cf.~\eq{eq:SMSRMu} and (\ref{eq:SMSRMD})), and hence it 
prefers high values of $M_{\rm LR}$ in order not to overshoot light neutrino masses (cf.~\eq{eq:SMSRnu}). 
Moreover, we note that the single-step breaking corresponds to the lower end of the 
axion mass window $m_a \approx 3 \times 10^{-7}$ GeV, which will be one 
of the first region to be tested, already in Phase 1 of ABRACADABRA (cf.~\fig{fig:maxionvsMLR}). 

In fact, a detailed fit of SM fermion masses and mixings within the minimal 
renormalizable Yukawa sector (cf.~the mass sum rules 
in \eqs{eq:SMSRMu}{eq:SMSRMe} and \eq{eq:SMSRnu}) 
could actually reveal a non-trivial constraint on the model and 
select an intermediate-scale value for $M_{\rm LR} \gg $ TeV. 
An independent argument for high-scale Left-Right breaking is given by thermal leptogenesis \cite{Fukugita:1986hr}, 
which in its simplest realization would suggest $M_{\rm LR} \gtrsim 10^9$ GeV (see e.g.~\cite{Davidson:2008bu}). 

Finally, in the presence of a strong first-order phase transition 
the Left-Right symmetry breaking 
can lead to the production of a stochastic gravitational wave background, 
that might leave its imprint on the gravitational wave spectrum 
of forthcoming space-based interferometers \cite{Caprini:2015zlo}.  
This can happen in some parameter space regions of the Left-Right symmetric scalar potential, 
resembling an approximate scale invariance \cite{Brdar:2019fur}. 
The latter work focussed on Left-Right breaking scales close to the TeV 
scale, but in principle detectable gravitational wave signals might arise also for 
$M_{\rm LR} \gg$ TeV. 
 
\subsubsection{Low-scale Left-Right breaking}

The Left-Right symmetry breaking scale can be as low as 
20 TeV (230 TeV) 
in the case where $\P$ is broken at the $M_{\rm LR}$ ($M_{\rm PS}$) 
scale (cf.~\Table{tab:twoloopscales}).\footnote{In this regime two-loop RG effects turn out to be very important 
(as it can be seen also from \fig{fig:MPSvsMLR}). 
For instance, in the case of $\P$ broken at the $M_{\rm LR}$ scale, the one-loop 
RG analysis yields $M_{\rm LR} \gtrsim 900$ TeV.} 
In particular, the lower bound is saturated in both cases for 
$M_{\rm PS} = 1.4 \times 10^{18}$ GeV, or equivalently (cf.~\eq{eq:MPSfa}) 
for a Pati-Salam/Peccei-Quinn breaking 
order parameter $v_{\S} = 3.4 \times 10^{18}$ GeV,  
that is of the order of the  
reduced Planck mass 
$M_{\rm Pl} / \sqrt{8 \pi}$. 
Although this numerical coincidence should not be taken too seriously, 
since it might be spoiled by scalar threshold effects, 
it is suggestive of a possible connection with gravity and it could be seen as 
a mild theoretical motivation for a low-scale Left-Right symmetry breaking 
scenario.\footnote{In the absence of the $\U(1)_{\rm PQ}$ 
symmetry the Black Hole Superradiance bound does not apply, 
and one can easily saturate present LHC direct/indirect limits. 
For instance, in the case of $\G_{\rm LR}$ we 
obtain that $M_{\rm LR} = 6$ TeV is obtained for 
$M_{\rm PS} = 2.5 \times 10^{18}$ GeV (cf.~also \fig{fig:MPSvsMLR}).}
In particular, the former case of $\P$ broken at $M_{\rm LR}$, 
corresponds to the most constrained version of the Left-Right 
symmetric model \cite{Senjanovic:1975rk,Mohapatra:1979ia}, 
whose phenomenology has beed studied in great detail 
in the recent years 
(see e.g.~\cite{Maiezza:2010ic,Tello:2010am,Bertolini:2014sua}). 
Although a $W_R$ mass of the order of 20 TeV is well beyond the 
direct/indirect $\approx 6$ TeV reach of the 
LHC \cite{Dev:2015kca,Ruiz:2017nip,Nemevsek:2018bbt,Chauhan:2018uuy}, 
flavour \cite{Bertolini:2014sua} and CP \cite{Maiezza:2014ala,Bertolini:2019out} 
violating observables offer sensitivities up to hundreds of TeV. 
A future 100 TeV hadron collider would be able instead to directly probe 
the lower end of the $M_{\rm LR}$ range, 
possibly in correlation with an axion signal at CASPEr-Electric for $m_a \gtrsim 10^{-11}$ eV.

\section{Conclusions}
\label{sec:concl}

In this work we have discussed the implementation of the PQ mechanism in a minimal realization
of the Pati-Salam (partial) unification scheme, where 
the axion mass is related to the Pati-Salam breaking scale. 
The latter was shown to be constrained by a RG analysis of  
(partial) gauge coupling unification. 
The main physics result is displayed in \fig{fig:maxionvsMLR}, which shows that the whole 
parameter space of the Pati-Salam axion will be probed in the late phases 
of the axion DM experiments ABRACADABRA and CASPEr-Electric. 
Possible correlated signatures connected with the breaking of the Left-Right symmetry group 
include future collider/flavour probes of a low-scale Left-Right 
breaking (as low as 20 TeV for a Pati-Salam breaking of the size of the reduced Planck mass -- 
cf.~\fig{fig:MPSvsMLR}) 
and, less generically, 
the imprint on the gravitational wave spectrum of the Left-Right phase transition.  
Other indirect constraints on the scale of Left-Right symmetry breaking 
might arise from a detailed fit of SM fermion masses and mixings within 
the minimal renormalizable 
Pati-Salam Yukawa sector 
or 
from a successful implementation of thermal leptogenesis. Both of them could be worth 
a future investigation, in order to further narrow down an axion mass range. 
Finally, some of the ingredients discussed in the present paper 
might serve as building blocks for a detailed investigation 
of what could arguably be considered 
the minimal $\SO(10) \times \U(1)_{\rm PQ}$ model, 
based on a $45_H + \bar{126}_H + 10_H$ reducible Higgs representation 
\cite{Bertolini:2009es,Bertolini:2012im}, with a complex adjoint hosting the 
SO(10) axion. 

While in the Introduction we have praised the nice aspects of the whole setup, 
here we would like to conclude with a more critical note. 
In the present formulation (as well as in all GUT$\times \U(1)_{\rm PQ}$ 
models known to the author and despite some earlier 
attempts of obtaining an automatic $\U(1)_{\rm PQ}$ from SU(9) \cite{Georgi:1981pu}) 
the PQ symmetry is imposed by hand, 
while it would be more satisfactory for it to arise as an accidental symmetry, possibly due to some 
underlying gauge dynamics.\footnote{Requiring a discrete 
$\Z_N$ gauge symmetry under which the 
axion multiplet transforms non-trivially appears to be just a technical, almost tautological solution.}  
Moreover, since global symmetries need not to be exact, 
it is unclear why the PQ symmetry should be an extremely good symmetry of UV physics, 
and in particular of quantum gravity, not to spoil the solution of the strong CP problem \cite{Kamionkowski:1992mf,Holman:1992us,Barr:1992qq}. 
This is particularly problematic for 
axion GUTs, since the issue gets worse in the $f_a \to M_{\rm Pl}$ limit. 
While we have nothing to say on this important problem, it would be desirable to have 
some fresh new ideas on how to tackle it (especially in GUTs). 
For the time being, we can pragmatically postpone the question until the discovery of the axion.

\section*{Acknowledgments} 
I wish to thank Enrico Nardi for a careful reading of the manuscript 
and for insightful observations. 
I also thank  
Stefano Bertolini,
Maurizio Giannotti, 
Luk\'a\v s Gr\'af, 
Ramona Gr\"ober,
Alberto Mariotti, 
Miha Nemev\v sek, 
Fabrizio Nesti, 
Michele Redi, 
Andreas Ringwald, 
Shaikh Saad,
Goran Senjanovi\'c 
and  
Carlos Tamarit 
for helpful and interesting discussions. 
This work is supported by the Marie Sk\l{}odowska-Curie 
Individual Fellowship grant AXIONRUSH (GA 840791).

\appendix

\section{Two-loop running and one-loop matching}
\label{app:2loopbf}

In this Appendix we collect the two-loop beta functions and the one-loop matching coefficient 
employed in the RG analysis. 

\subsubsection*{Two-loop beta functions} 

Let us denote the product of gauge factors $G= G_1 \times \ldots \times G_N$.
The two-loop RG equations for the corresponding gauge couplings $g_i$ ($i=1,\ldots, N$) can be written as 
\begin{equation}
\label{alpha2loops}
\mu \frac{d}{d\mu}\alpha^{-1}_{i}=-\frac{a_{i}}{2\pi}
-\sum_j \frac{b_{ij}}{8\pi^2}\alpha_{j} \, ,
\end{equation}
where $\alpha_i = g_i^2 / (4\pi)$.
The one- and two-loop beta coefficients in the $\overline{\rm MS}$ scheme 
are \cite{Machacek:1983tz} 
(no summation over $i$) 
\begin{align}
\label{oneloopbf}
a_{i}&=- \frac{11}{3} C_2(G_i) + \frac{4}{3} \sum_F \kappa S_2(F_i) + \frac{1}{3} \sum_S \eta S_2(S_i)\,, \\
\label{twoloopbf}
b_{ij}&= 
\left[- \frac{34}{3} \left( C_2(G_i) \right)^2 
+  \sum_F \left( 4 C_2(F_i) + \frac{20}{3} C_2(G_i) \right) \kappa S_2(F_i) \right. \nonumber \\
& \left.  + \sum_S \left( 4 C_2(S_i) + \frac{2}{3} C_2(G_i) \right) \eta S_2(S_i) \right]\delta_{ij} \nonumber \\
&+ 4 \Big[  \sum_F \kappa C_2(F_j) S_2(F_i) + \sum_S \eta C_2(S_j) S_2(S_i)  \Big] \, , 
\end{align}
where $G_i$ denotes the $i$-th gauge factor, 
$S_{2} (R_i)= T(R_i) d(R)/d(R_i)$ in terms of the Dynkin index 
(with normalization $\frac{1}{2}$ for the fundamental)
of the
representation $R_i$, $T(R_i)$, and the multiplicity factor $d(R)/d(R_i)$, 
with $d(R_i)$ ($d(R)$) denoting the dimension of the representation under $G_i$ ($G$). 
The latter are related to the Casimir invariant, 
$C_2(R_i)$, via $C_2 (R_i) d(R_i) = T(R_i) d(G_i)$, 
where $d(G_i)$ is the dimension of $G_i$. 
$\kappa=1,\frac{1}{2}$ for Dirac, Weyl fermions ($F$) 
and
$\eta=1, \frac{1}{2}$ for complex, real scalar ($S$) fields, respectively. 
The Yukawa contribution in the two-loop beta coefficient has been 
neglected.\footnote{The effects of the Yukawa couplings can be at leading order approximated 
by constant negative shifts of the one-loop gauge beta coefficients, $a_i \to a_i - \Delta a_i$, 
with $\Delta a_i \lesssim 1\%$ \cite{Bertolini:2009qj}. For instance, in the case of SO(10) this resulted 
in relative variations on the unified gauge coupling and the unification scale 
at the level of $1\permil$ and $1\%$, respectively \cite{Bertolini:2009qj}.} 

Specifically, for the two running stages considered in this work we have 
(using the ESH intermediate-scale scalars in \Table{tab:ESH}): 
\begin{itemize}
\item $\G_{\rm SM} \equiv \SU(3)_{C} \times \SU(2)_L \times \U(1)_Y$ ($M_Z \to M_{\rm LR}$ running)
\beq 
\label{eq:bfSM}
a^{\rm SM} = \(-7,-\tfrac{19}{6},\tfrac{41}{10}\) \, , \qquad 
b^{\rm SM} = 
\left(
\begin{array}{ccc}
 -26 & \frac{9}{2} & \frac{11}{10} \\
 12 & \frac{35}{6} & \frac{9}{10} \\
 \frac{44}{5} & \frac{27}{10} & \frac{199}{50} \\
\end{array}
\right) \, .
\eeq
\item $\G_{\rm LR} \equiv \SU(3)_{C} \times \SU(2)_L \times \SU(2)_R \times \U(1)_{B-L} \times \P$ ($M_{\rm LR} \to M_{\rm PS}$ running)
\beq 
\label{eq:bfLRP}
a^{\rm LR} = \(-7,-\tfrac{7}{3},-\tfrac{7}{3},7\) \, , \qquad 
b^{\rm LR} = 
\left(
\begin{array}{cccc}
 -26 & \frac{9}{2} & \frac{9}{2} & \frac{11}{2} \\
 12 & \frac{80}{3} & 3 & \frac{297}{2} \\
 12 & 3 & \frac{80}{3} & \frac{297}{2} \\
 4 & \frac{81}{2} & \frac{81}{2} & \frac{1265}{2} \\
\end{array} 
\right) \, .
\eeq
\item $\G^{\slashed{\P}}_{\rm LR} \equiv \SU(3)_{C} \times \SU(2)_L \times \SU(2)_R \times \U(1)_{B-L}$ ($M_{\rm LR} \to M_{\rm PS}$ running)
\beq 
\label{eq:bfLRNoP}
a^{\rm LR} = \(-7,-3,-\tfrac{7}{3},\tfrac{11}{2}\) \, , \qquad 
b^{\rm LR} = \left(
\begin{array}{cccc}
 -26 & \frac{9}{2} & \frac{9}{2} & \frac{11}{2} \\
 12 & 8 & 3 & \frac{33}{2} \\
 12 & 3 & \frac{80}{3} & \frac{297}{2} \\
 4 & \frac{9}{2} & \frac{81}{2} & \frac{671}{2} \\
\end{array}
\right) \, .
\eeq
where we considered both the case in which $\P$ is broken at the $M_{\rm LR}$ scale ($\G_{\rm LR}$) 
and at the $M_{\rm PS}$ scale ($\G^{\slashed{\P}}_{\rm LR}$) 
\end{itemize}

\subsubsection*{One-loop matching coefficients} 

The general form of the one-loop matching condition between effective theories in the framework of 
dimensional regularization was derived in \cite{Weinberg:1980wa,Hall:1980kf} 
(see also \cite{Bertolini:2009qj} for the inclusion of $\U(1)$ mixing). 
Considering for definiteness the case of a simple group $G$ spontaneously broken into subgroups $G_i$, 
the one-loop matching (at the matching scale $\mu$) for the gauge couplings can be written as \cite{Bertolini:2013vta}
\beq 
\label{eq:1loopmatchgen}
\alpha^{-1}_i (\mu) = \alpha^{-1} (\mu) - 4\pi \lambda^{i}  (\mu) \, ,
\eeq
where 
\begin{align}
\label{eq:1loopmatchgen2}
\lambda^{i}  (\mu) &= \frac{1}{12 \pi} ( C_2(G) - C_2(G_i) ) \nonumber \\
&+ \frac{1}{2\pi} 
\[ 
-\frac{11}{3} \Tr T_{V_i}^2 \log \frac{M_{V_i}}{\mu} 
+\frac{4}{3} \kappa  \Tr T_{F_i}^2 \log \frac{M_{F_i}}{\mu} 
+\frac{1}{3} \eta  \Tr T_{S_i}^2 \log \frac{M_{S_i}}{\mu} 
\]
\, ,
\end{align}
with $V$, $F$ and $S$ denoting the massive vectors, 
fermions and scalars that are integrated out at the matching scale $\mu$. 
Note that differently from \cite{Weinberg:1980wa,Hall:1980kf} 
the (Feynman gauge) Goldstone bosons have been conveniently 
included in the scalar part of the expression, so that the matching coefficients 
resembles the structure of the one-loop beta coefficients in \eq{oneloopbf}.

Specifically, for the two matching scales considered in this work we have: 
\begin{itemize} 
\item $M_{\rm LR}$ ($\G_{\rm SM} \leftrightarrow \G^{(\slashed{\P})}_{\rm LR}$ matching) 
\begin{align}
\label{1lalpha2LR}
\alpha_1^{-1} (M_{\rm LR}) &= \frac{3}{5} \(  \alpha_R^{-1} (M_{\rm LR}) - \frac{1}{6\pi} \)
+ \frac{2}{5} \alpha_{B-L}^{-1} (M_{\rm LR}) \, , \\
\label{1lalpha2LR}
\alpha_L (M_{\rm LR}) &= \alpha_L (M_{\rm LR}) \, , \\
\label{1lalpha3LR}
\alpha_C (M_{\rm LR}) &= \alpha_{C} (M_{\rm LR}) \, .  
\end{align}
\item $M_{\rm PS}$ ($\G^{(\slashed{\P})}_{\rm LR} \leftrightarrow \G_{\rm PS}$ matching) 
\begin{align}
\label{1lalpha1PS}
\alpha^{-1}_{B-L} (M_{\rm PS}) &= \alpha_{C}^{-1} (M_{\rm PS}) - \frac{1}{4\pi} = \alpha_{\rm PS}^{-1} (M_{\rm PS}) 
- \frac{1}{3\pi}  \, , \\
\label{1lalpha2PS}
\alpha_R (M_{\rm PS}) &= \alpha_L (M_{\rm PS}) = \alpha_{\rm LR} (M_{\rm PS}) \, .
\end{align}
\end{itemize}

\bibliographystyle{utphys.bst}
\bibliography{bibliography}

\end{document}